\documentclass[a4paper,11pt]{article}

\usepackage{jheppub} 

\usepackage[normalem]{ulem}  

\usepackage{color}
\input{colordvi.tex}
\usepackage{graphicx}
\usepackage{float}
\usepackage{wrapfig}
\usepackage{bm}
\usepackage{amsmath,amssymb,ascmac}
\usepackage{subfigure}
\usepackage{verbatim}
\usepackage{makeidx}

\newcommand{\Pcal}{{\mathcal{P}}}
\newcommand{\hOcal}{{\hat{\mathcal{O}}}}
\newcommand{\with}{{\quad\mathrm{with}}\quad}
\newcommand{\msr}{{\mathrm{MSR}}}
\newcommand{\hx}{\hat{x}}
\newcommand{\hp}{\hat{p}}
\newcommand{\average}[1]{\langle#1\rangle_{\mathrm{eq}}}

\usepackage{booktabs}

\title{
Effective field theory of time-translational symmetry breaking  
in nonequilibrium open system
}

\author[a]{Masaru Hongo,} 
\author[b]{Suro Kim,}
\author[b]{Toshifumi Noumi,}
\author[c]{and Atsuhisa Ota}

\affiliation[a]{iTHEMS Program, RIKEN, Wako 351-0198, Japan}
\affiliation[b]{Department of Physics, Kobe University, Kobe 657-8501, Japan}
\affiliation[c]{
Institute for Theoretical Physics and Center for Extreme Matter and Emergent Phenomena,
Utrecht University, Princetonplein 5, 3584 CC Utrecht, The Netherlands}

\emailAdd{masaru.hongo@riken.jp}
\emailAdd{s-kim@stu.kobe-u.ac.jp}
\emailAdd{tnoumi@phys.sci.kobe-u.ac.jp}
\emailAdd{a.ota@uu.nl}

\preprint{KOBE-COSMO-18-05}

\abstract{
We develop the effective field theoretical (EFT) approach to time-translational symmetry breaking
of nonequilibrium open systems based on the Schwinger-Keldysh formalism. In the Schwinger-Keldysh formalism, all the symmetries of the microscopic Lagrangian are doubled essentially because the dynamical fields are doubled to describe the time-evolution along the closed-time-path. The effective Lagrangian for open systems are then obtained by coarse-graining the microscopic Schwinger-Keldysh Lagrangian. As a consequence of coarse-graining procedure, there appear the noise and dissipation effects, which explicitly break the doubled time-translational symmetries into a diagonal one. We therefore need to incorporate this symmetry structure to construct the EFT for Nambu-Goldstone bosons in symmetry broken phases of open systems. Based on this observation together with the consistency of the Schwinger-Keldysh action, we construct and study the general EFT for time-translational symmetry breaking in particular, having in mind applications to synchronization, time crystal, and cosmic inflation.
}

\begin{document} 
\maketitle
\flushbottom

\section{Introduction}

The effective field theory (EFT) approach based on the symmetry structure 
provides a universal framework for the low-energy dynamics~
\cite{Manohar:1996cq}. 
It is very powerful especially when applied to the dynamics 
in the symmetry broken phase, whose low-energy dynamics 
is controlled by the massless Nambu-Goldstone (NG) modes 
associated with spontaneously broken symmetries.
Although it has been originally developed in the context of 
high-energy particle physic~\cite{Coleman:1969sm,Callan:1969sn,Weinberg:1978kz},
its range of applications covers  almost all areas of physics
from condensed matters and particle physics to cosmology. 
Moreover, recent applications may reach beyond traditional subjects of  
physics, e.g., to active matters~\cite{TonerTu1,TonerTu2} 
such as schools of fish and flocks of birds!

\medskip
Among various applications of the EFT approach, one interesting direction recently explored intensively is the application to real-time nonequilibrium dynamics of open systems, where the dynamics in interests is affected by the noise and dissipation originated from environments. 
The most popular example is dissipative (fluctuating) hydrodynamics, 
where the effect of both dissipation and fluctuation take place~\cite{Grozdanov:2013dba,Haehl:2015uoc,Crossley:2015evo,Glorioso:2017fpd,Jensen:2017kzi,Haehl:2018lcu,Jensen:2018hse}. 
In dissipative hydrodynamics, 
we only focus on the conserved charge densities and associated hydrodynamic 
modes, which means that other all high-energy modes play a role of environments.
Furthermore, by incorporating violation of conservation law into the hydrodynamic equations, we may describe collective behaviors of active matter
(See Ref.~\cite{RevModPhys.85.1143} and references therein for a review). 
Another typical example is cosmology: 
In cosmology, we cannot observe degrees of freedom outside the cosmological horizon, which play the role of environment when discussing the dynamics inside the horizon. 
Also, in inflationary scenarios based on high energy theories, inflaton and graviton are generically coupled to massive fields, which we cannot probe directly~\cite{Chen:2009zp,Baumann:2011nk,Noumi:2012vr,Arkani-Hamed:2015bza}. Such a hidden sector again plays the role of environment.  
As we can see from these broad examples, 
the viewpoint of open systems is universal among various systems in nature. 
Nevertheless, it is just recent that we start to focus on the 
nonequilibrium open systems from the well-established EFT viewpoint.

\medskip
In this paper we develop the EFT approach to symmetry broken phases of open systems. 
In particular, we focus on the time-translational symmetry breaking, 
having two concrete applications in mind. 
One is the synchronization phenomena~\cite{Kuramoto,MoriKuramoto} 
taking place in e.g. reaction-diffusion systems, or a kind of time crystal 
in condensed matter systems~\cite{PhysRevLett.96.230602,PhysRevLett.99.140402,sieberer2016keldysh}\footnote{
Contrary to the original na\"ive proposal of the time crystal~\cite{Wilczek:2012jt}, 
the realization of the quantum time crystal in the ground state or 
thermal equilibrium is theoretically rejected~\cite{Bruno:2013mva,Watanabe:2014hea,Yamamoto:2015fxa}.
Nevertheless, as we discuss in this paper, if we extend the notion of 
spontaneous symmetry breaking in the excited or non-equilibrium state, 
there is still a possibility to realize it.
See also~\cite{Buca:2018ccq}.
}.
In these systems, we have a time-periodic physical observable, 
which means our continuous time-translational symmetry is broken to 
a discrete one. 
The other is the cosmic inflation~\cite{Starobinsky:1980te,Guth:1980zm,Sato:1980yn}, 
where the time-dependent inflaton background breaks the time-translational 
symmetry and the corresponding NG boson sources the structure 
in our universe 
such as temperature fluctuations of the 
Cosmic Microwave Background (see, e.g.,~\cite{Baumann:2009ds,Senatore:2016aui} for review articles).
The same symmetry breaking pattern also appears in the phenomenological 
approach to dark energy called the quintessence~\cite{Gubitosi:2012hu}.
For isolated systems, the EFT approach for such a time-translational symmetry breaking is already well studied in the context of cosmic inflation~\cite{Cheung:2007st}. By applying recent developments in the EFT of dissipative fluids~\cite{Crossley:2015evo,Glorioso:2017fpd}, we further incorporate effects of noise and dissipations in open systems into this EFT framework for time-translational symmetry breaking.

\medskip
The organization of the paper is as follows. 
In Sec.~\ref{section:TT},
we first introduce basic concepts such as the doubled time-translational symmetry
and its breaking attached to open systems by using a simplest model---the Brownian particle in the presence of the dissipation and random force. 
In Sec.~\ref{section:ConstructingEFT}, we then construct 
the EFT of open systems with time-translational symmetry breaking 
based on the Schwinger-Keldysh formalism and its symmetry structure.
By using the constructed effective Lagrangian, 
we derive the dispersion relations of the Nambu-Goldstone mode.
In addition to the general consequence, 
we also demonstrate a simple UV model composed of 
a single scaler field which can be explicitly analyzed, 
and show the relation between the low-energy (Wilson) coefficient in our EFT  
and parameters in the UV theory.
Sec.~\ref{section:Summary} is devoted to the summary and outlook.

\section{Time-translational symmetry of Brownian motion}
\label{section:TT} 

First of all, we elaborate the symmetry structure 
attached to open systems by analyzing a simple example; 
that is, the Brownian particle system. 
In Sec.~\ref{section:ClassicalBrownian}, 
starting from the Langevin equation,
we introduce the Martin-Siggia-Rose (MSR) 
path-integral formalism and the Fokker-Planck formalism. 
In Sec.~\ref{section:BrownianSymmetries}, based on these two formalisms, 
we show that there are two kinds of generators (charges) related to 
two time-translational symmetries, and provide our \textit{weak} criterion for 
spontaneous symmetry breaking (SSB) of time-translational symmetry 
in open systems. 
We also explain how the fluctuation dissipation relation can be incorporated 
as a consequence of a discrete symmetry. 
In Sec.~\ref{micro:origin}, we discuss the origin of the doubled symmetries 
and the discrete symmetry from the viewpoint of underlying quantum theory 
based on the Schwinger-Keldysh formalism.
All materials given in this section serve as a basis for the subsequent section.
Readers familiar with the MSR formalism and the Schwinger-Keldysh 
 formalism can skip most of this section after checking 
our \textit{weak} definition of SSB provided in the end of Sec.~\ref{sec:TimeTrSym}
.

\subsection{Three equivalent formalisms}
\label{section:ClassicalBrownian}
The simplest intuitive way to describe stochastic systems
with the fluctuation-dissipation (e.g. Brownian particles in a fluid) is 
to use \textit{the Langevin equation}.
Nevertheless, the Langevin equation is not so useful to discuss 
symmetry inherent in the stochastic system, and it is more helpful to 
employ other formulations equivalent to that.
Thus, starting from the Langevin equation for the Brownian particle, 
we introduce two equivalent formalisms for classical stochastic systems: 
\textit{the Fokker-Planck formalism} and 
\textit{the Martin-Siggia-Rose~(MSR) formalism}, 
which, based on the analogy to quantum theory, 
correspond to the canonical operator formalism 
and path-integral formalism, respectively.  

\subsubsection{Langevin equation}
\label{section:Langevin}
To illustrate the symmetry structure and associated conserved charges 
in open systems, let us consider the Brownian motion described 
by the underdamped Langevin equation:
\begin{align}
 M \ddot X_R (t) =- V'\big(X_R(t)\big) - M \gamma \dot X_R(t) +\xi (t) \,,
 \label{eq:langevin1}
\end{align}
where $X_R(t)$ denotes the position of the Brownian particle with a mass $M$,
$V'(X_R) = \dfrac{\partial V}{\partial X_R}$ with a
 potential energy $V(X_R)$, 
$\gamma$ a damping coefficient, 
and $\xi (t)$ a random force acting on the Brownian particle. 
The random force $\xi(t)$ 
is given by the Gaussian white noise:
\begin{equation} 
 \langle \xi(t) \rangle_\xi  = 0 , \quad 
  \langle \xi(t)\xi(t')\rangle_\xi 
  = A \,\delta(t-t')\,,
  \label{eq:noise}
\end{equation}
where we introduced the noise average $\langle \mathcal{O}(\xi) \rangle_\xi$ as 
\begin{equation}
 \langle \mathcal{O}(\xi) \rangle_\xi 
  \equiv \int \mathcal{D} \xi
  \mathcal P[\xi] \mathcal{O} (\xi) 
  \quad \mathrm{with} \quad 
  \mathcal P[\xi] \equiv \exp \left[ - \frac{1}{2A} \int dt\, \xi^2(t) \right].
  \label{eq:NoiseAvg}
\end{equation}
When we have the fluctuation-dissipation 
relation (FDR), it implies that 
$A= 2 M \gamma T$ 
with the temperature of a surrounding fluid in thermal equilibrium $T$.
However, considering more general situations, 
we leave $A$ a general constant in the following discussion. 
Since Eqs.~\eqref{eq:langevin1} and~\eqref{eq:noise} do not 
have explicit time dependence, 
this system enjoys the time-translational symmetry, in other words, Eq.~\eqref{eq:langevin1} is covariant under $X_R(t)\to X'_R(t)=X_R(t+\epsilon)$ and $\xi(t)\to \xi'(t)=\xi(t+\epsilon)$.
Then, a natural question arises here: 
\textit{``What is the corresponding conserved charge associated with 
this time-translational symmetry?''}
Obviously, it is not the energy of the Brownian particle, 
which does not conserve due to the existence of the noise and dissipation:
\begin{align}
 \frac{d}{dt}\left[\frac{1}{2} M \dot X^2_R + V \big(X_R \big) \right] 
 = - M \gamma \dot X^2_R + \dot X_R \xi \neq 0 . \,\label{consviolgv}
\end{align}
It is also instructive to demonstrate that this energy of the 
Brownian particle does not generate time-translations. 
To demonstrate this, 
let us introduce the momentum $P_R$ and rewrite Eq.~\eqref{eq:langevin1} as
\begin{align}
 \label{eq:eom_xp}
 M \dot{X}_R  = P_R ,  \quad
 \dot{P}_R = -V' (X_R) - \gamma P_R + \xi \,.
\end{align}
If we introduce the energy $H_R$ of the Brownian particle by
\begin{align}
 H_R = \frac{P^2_R}{2M} + V(X_R) \,,
\end{align}
the equations of motion can be rewritten as
\begin{align}
 \dot{X}_R=\frac{\partial H_R}{\partial P_R} 
 \,,
 \quad
 \dot{P}_R=-\frac{\partial H_R}{\partial X_R}- \gamma P_R + \xi\, ,
 \label{eq:langevin2}
\end{align}
so that $\dot{P}_R \neq \{P_R, H_R \}_{PB}$. 
Here $\{A,B\}_{PB}$ is the Poisson bracket.
If there are no noise and dissipation, 
we may identify $(X_R,P_R)$ as a pair of canonical variables and 
the particle energy $H_R$ as the Hamiltonian which properly 
generates time-translation of our system. 
However, it is not the case in dissipative systems. 
This is the main obstruction to describing Brownian motion 
by a single pair of canonical variables. 
Also, in this formulation, it is not yet clear what is 
the conserved charge associated to the time-translational symmetry.

\subsubsection{Martin-Siggia-Rose formalism}
\label{section:lmsrfp}
To identify the generator for the above time-translational symmetry, 
it is convenient to employ other equivalent formalisms: 
the Martin-Siggia-Rose (MSR) path-integral formalism~\cite{PhysRevA.8.423,Janssen1976,PhysRevB.18.4913} and the Fokker-Planck (operator) formalism. 
Starting from the Langevin equation, 
we first introduce the MSR formalism.

As is the case for the usual path-integral formalism for quantum theory, 
what we focus on in the MSR formalism is a set of correlation functions;
that is, we are interested in
\begin{equation}
 \langle \mathcal{O}[X_R,P_R] \rangle_\xi 
  \equiv \int \mathcal{D} \xi \mathcal P[\xi] \mathcal{O}[X_\xi,P_\xi], 
  \label{eq:correlation}
\end{equation}
where $\{X_\xi(t_n), P_\xi(t_n)\}$ 
denotes a set of solutions of the Langevin equation 
\eqref{eq:langevin2}, and $\mathcal P[\xi]$ is defined in Eq.~(\ref{eq:NoiseAvg}).
If we choose e.g. 
$\mathcal{O}[X_R,P_R] = X_R(t_1) X_R(t_2) \cdots X_R(t_n)$, this provides 
the $n$-point function of the position of the Brownian particle, 
which has basic information on the stochastic processes under consideration.
Then, inserting the identity 
$1 = \displaystyle{\int} \mathcal{D} X_R \mathcal{D} P_R
 \delta (X_R - X_\xi ) \delta (P_R - P_\xi )$ and 
replacing $\{X_\xi (t_n), P_\xi(t_n)\}$ with the integration variable 
$\{X_R(t_n), P_R(t_n)\}$, we can calculate Eq.~\eqref{eq:correlation} 
in the following way:
\begin{equation}
 \begin{split}
  \langle \mathcal{O}[X_R ,P_R] \rangle_\xi
  &= \int \mathcal{D} X_R  \mathcal{D} P_R \mathcal{D} \xi 
  \delta (X_R - X_\xi)  \delta (P_R - P_\xi) \mathcal P[\xi]
  \mathcal{O}[X_R,P_R] \\
  &= \int \mathcal{D} X_R \mathcal{D} P_R  J 
  \big\langle \delta (\mathrm{EoM}_{X_\xi}) \delta (\mathrm{EoM}_{P_\xi} )
  \big\rangle_\xi 
  \mathcal{O}[X_R,P_R] .
  \label{eq:2.9}
   \end{split}
\end{equation}
Here we rewrite the delta functional in terms of the 
equations of motion $\mathrm{EoM}_{X_\xi, P_\xi}$ defined by
\begin{equation}
 \mathrm{EoM}_{X_\xi}
 \equiv 
   \dot{X}_R - \frac{P_R}{M} , \quad
   \mathrm{EoM}_{P_\xi}
 \equiv 
  - \dot{P}_R + \big[ - V'(X_R) - \gamma P_R + \xi \big],
\end{equation}
which brings about appearance of the Jacobian 
$J \equiv \det \left| \dfrac{\delta (\mathrm{EoM}_{X_\xi})}{\delta X_R} \right|
\det \left| \dfrac{\delta (\mathrm{EoM}_{P_\xi})}{\delta P_R} \right|$.
In general, this Jacobian may play an important role when we perform 
e.g. loop calculations. 
However, since each equation of motion does not contain terms nonlinear
in $X_R$ and $P_R$, respectively,
we don't have to consider the Jacobian part.
By using the Fourier representation of the delta functional,
\begin{align}
 \delta(\mathrm{EoM}_{X_\xi})
 &\propto
 \int \mathcal{D}P_A 
 \exp\left[ i\int dt P_A \big( \mathrm{EoM}_{X_\xi} \big) \right],\\
 \delta(\mathrm{EoM}_{P_\xi})
 &\propto \int \mathcal{D} X_A 
 \exp \left[ i\int dt X_A \big( \mathrm{EoM}_{P_\xi} \big) \right] ,
\end{align}
we can perform the Gaussian noise integral, which results in 
\begin{equation}
 \begin{split}
  \langle \mathcal{O}[X_R,P_R] \rangle_\xi
  &= \mathcal N \int \mathcal{D} X_R \mathcal{D} X_A \mathcal{D} P_R  \mathcal{D} P_A J
  \exp \Big( iS_{\mathrm{MSR}} [X_R,X_A,P_R,P_A] \Big)
  \mathcal{O}[X_R,P_R]
 \end{split}
\end{equation}
with an overall constant $\mathcal N$,
where we introduced the phase-space MSR effective action, 
\begin{align}
 iS_{\mathrm{MSR}}[X_R,X_A,P_R,P_A] 
 &=\int dt\, \left[  i P_A \left( \dot{X}_R - \frac{P_R}{M} \right) 
 - i X_A \big(\dot P_R+ V'(X_R) + \gamma P_R  \big)
 - \frac{A}{2}X_A^2\right].
 \label{eq:MSR1}
 \end{align}
Furthermore, 
integrating out momenta $P_R$ and $P_A$, 
one finds the configuration-space MSR effective action 
for the Brownian particle:
\begin{align}
 iS_{\rm MSR}[X_{R},X_{A}]=\int dt\, 
 \Big[  -i X_A \big( M\ddot{X}_R + M \gamma\dot{X}_{R} + V'(X_{R})
  \big)  - \frac{A}{2}X_A^2 \Big]. \label{eq:MSR2}
\end{align}
Thus, our classical stochastic system---the Brownian motion---is written 
in terms of the path integral over the original variable $X_R(t)$ and 
the auxiliary variable $X_A(t)$ known as the response 
field
or the advanced variable.
Since we now have the time-translationally invariant action,
it is straightforward to discuss the symmetry property 
associated with the time-translational symmetry. 
But before moving to the discussion on symmetry, 
we introduce another useful formalism---the Fokker-Planck formalism---in 
the next subsection.

\subsubsection{Fokker-Planck formalism}
We next move on to the corresponding operator formalism
called the Fokker-Planck formalism
to discuss the symmetry property.
In this formalism, the probability distribution function 
$\Pcal[X_R,P_R;t]$ defined by the following equation
plays a central role:
\begin{equation}
 \Pcal[X_R,P_R;t] 
  \equiv 
  \big\langle \delta \big(X_R - X_\xi (t) \big) 
  \delta \big( P_R - P_\xi(t)  \big) \big\rangle_\xi,
\end{equation}
where the stochastic variables $\{X_\xi (t),P_\xi (t)\}$ obey the Langevin equation \eqref{eq:langevin2}.
Then, carefully calculating the time derivative of this probability 
distribution by using the Langevin equation, 
we obtain the following Fokker-Planck equation%
\footnote{
To be precise, this is not the equation known as the Fokker-Planck equation 
but known as the Kramers equation since we are considering the 
underdamped Langevin equation.
}:
\begin{equation}
 \begin{split}
  \frac{\partial \Pcal}{\partial t} 
  = \left[ -\frac{P_R}{M} \frac{\partial}{\partial X_R} 
  + \frac{\partial}{\partial P_R} 
  \left( \frac{\partial V}{\partial X_R}  + \gamma P_R \right)
  + \frac{\partial^2}{\partial P^2_R} \frac{A}{2} \right] 
  \Pcal .
 \end{split}
 \label{eq:FP1}
\end{equation}
In order to clarify the analogy to the operator formalism 
of quantum theory, we define the following set of operators
\begin{equation}
 \hat{X}_R \equiv X_R, \quad 
  \hat{P}_A \equiv \frac{1}{i}\dfrac{\partial}{\partial X_R}
  \quad \mathrm{and} \quad
  \hat{P}_R \equiv P_R , \quad 
  \hat{X}_A  \equiv i \dfrac{\partial}{\partial P_R} ,
\end{equation}
where the former and latter two operators form a set of conjugate variables 
and thus satisfy
\begin{align}
 [\hat{X}_R, \hat{P}_A] = [\hat{X}_A, \hat{P}_R] = i,  
 \quad 
(\mathrm{others}) = 0  \, .
\end{align}
In the same way as quantum theory, 
square brackets denotes the commutation relation: 
$[\hat{A}, \hat{B}] \equiv \hat{A} \hat{B} - \hat{B} \hat{A}$.
With the help of these operators, 
we can rewrite the Fokker-Planck equation \eqref{eq:FP1} in 
the Schr\"odinger-like equation:
\begin{equation}
 \partial_t \Pcal [X_R,P_R;t] = - \hat{H}_A \Pcal [X_R,P_R;t]\,,\label{eq:sch}
\end{equation}
where we introduced the Fokker-Planck Hamiltonian $\hat{H}_A$ as%
\footnote{
Alternatively, we can directly derive this Fokker-Planck Hamiltonian 
from the MSR formalism by using the Legendre transformation.
} 
\begin{equation}
 \begin{split}
  \hat{H}_A 
  = \frac{i}{M} \hat{P}_R \hat{P}_A  
  + i \hat{X}_A V'(\hat{X}_R) 
  + i\gamma \hat{X}_A \hat{P}_R
  + \frac{A}{2} \hat{X}_A^2.
 \end{split}
 \label{eq:FP2}
\end{equation}
Here we neglect $P_R$ dependence of the noise amplitude $A$ 
and the damping coefficient $\gamma$.
Since we have the Schr\"odinger-like equation for 
the probability distribution function, 
we can think of the Fokker-Planck Hamiltonian as the generator of the 
time-translation.
In fact, we can employ the Heisenberg-like picture to describe 
the time evolution of an arbitrary operator $\hOcal (t)$ as 
(See e.g. Ref.~\cite{Zinn1996quantum}) 
\begin{equation}
 \partial_t \hOcal (t) = [\hat{H}_A, \hOcal (t)]
  \with
  \hOcal (t) = e^{ \hat{H}_A t } \hOcal e^{- \hat{H}_A t}.
  \label{eq:EoMinFP}
\end{equation}
Let us confirm that the Fokker-Planck Hamiltonian properly acts on 
our physical operators $\{\hat{X}_{R,A}, \hat{P}_{R,A}\}$ 
as the generator of time-translation:
\begin{align}
 \partial_t \hat{X}_R &= [\hat{H}_A, \hat{X}_R] = \frac{\hat{P}_R}{M} 
 \,\label{HPtrsf:1},
 \\
 \partial_t \hat{P}_R &= [\hat{H}_A, \hat{P}_R] 
 = -V'(\hat{X}_R) - \gamma \hat{P}_R + iA \hat{X}_A \,, 
 \\ 
 \partial_t \hat{X}_A &= [\hat{H}_A, \hat{X}_A] 
 = \frac{\hat{P}_A}{M} + \gamma \hat{X}_A\,,
 \\
 \partial_t \hat{P}_A &= [\hat{H}_A, \hat{P}_A] 
 = - \hat{X}_A  V''(\hat{X}_R) \,\label{HPtrsf:4},
\end{align}
which reproduce the expected equations of motion following from 
the MSR action \eqref{eq:MSR1}.

\subsection{Symmetry associated with the Brownian motion}
\label{section:BrownianSymmetries}

With the help of the MSR and Fokker-Planck formalisms introduced in the previous 
subsection, we demonstrate symmetries and generators attached to 
the Brownian motion: 
One is explicitly broken when there are fluctuation and dissipation, 
and the other gives an answer to our question raised 
in Sec.~\ref{section:Langevin}.
After presenting our \textit{weak} definition of SSB of 
time-translational symmetry for open systems, we also discuss 
another discrete symmetry which contains information on 
the fluctuation-dissipation relation.

\subsubsection{Time-translational symmetries}
\label{sec:TimeTrSym}
Let us look at the symmetry structure of the Brownian particle system 
from the Fokker-Planck and MSR viewpoint. 
First of all, as is already mentioned, the MSR effective action \eqref{eq:MSR2} 
does not have explicit time dependence, hence it is invariant under the following time-translation labelled by $\epsilon_R$:
\begin{equation}
 \begin{cases}
  X_R (t) \to X_R' (t) = X_R ( t + \epsilon_R ), \\ 
  X_A (t) \to X_A' (t) = X_A ( t + \epsilon_R ).
 \end{cases}
 \label{eq:Timetr1}
\end{equation}
This invariance provides us a conserved charge given by
\begin{equation}
  H_A 
  = i M\dot{X}_R \dot{X}_A + i X_A V'(X_R) + \frac{A}{2} X_A^2 ,
\end{equation}
which is nothing but the Fokker-Planck Hamiltonian expressed 
in the configuration space.
Recalling Eq.~\eqref{eq:EoMinFP}, 
we can easily understand this in the Fokker-Planck formalism since 
the time evolution of operators is generated by $\hat{H}_A$, 
and thus $\hat{H}_A$ is trivially conserved:
\begin{align}
\frac{d }{dt}\hat H_{A}=[\hat H_{A},\hat H_{A}]=0.
\end{align}
Hence, the Fokker-Planck Hamiltonian $\hat H_{A}$ is 
the conserved charge, or the generator in other words, of time-translation 
which we have been looking for.

\medskip
Then, what about our original Hamiltonian $H_R$?
By the use of Eqs.~(\ref{HPtrsf:1}) to (\ref{HPtrsf:4}) 
in the Fokker-Planck formalism, 
we can clarify properties of the original Hamiltonian for the Brownian motion:
\begin{align}
 \hat{H}_R=\frac{\hat{P}^2_R}{2M} + V(\hat{X}_R)\,.
\end{align}
As we mentioned, it does not conserve due to the existence of 
the fluctuation and dissipation. 
In fact, the Fokker-Planck Hamiltonian enables us to find 
\begin{align}
 \frac{d}{dt} \hat{H}_R =  [\hat{H}_A, \hat{H}_R] 
 = -\frac{\gamma}{M} \hat{P}^2_R 
 + i\frac{A}{2M} 
 \big(\hat{X}_A \hat{P}_R + \hat{P}_R \hat{X}_A \big) \,,
\end{align}
which vanishes only if there are no noise and dissipation, i.e., $\gamma=A=0$. 
Defining an infinitesimal transformation generated by $\hat H_R$ as 
$\delta_{\epsilon_A} \hat{\mathcal{O}} 
\equiv i\epsilon_A [\hat{H}_R, \hat{\mathcal{O}}]$, 
we also obtain
\begin{align}
 \label{Hx}
 \delta_{\epsilon_A} \hat{X}_R &= i\epsilon_A[\hat{H}_R, \hat{X}_R] = 0\,,
 \\
 \delta_{\epsilon_A} \hat{P}_R &= i\epsilon_A [\hat{H}_R, \hat{P}_R] = 0\,,
 \\
 \label{Hx_A}
 \delta_{\epsilon_A} \hat{X}_A &= i\epsilon_A [\hat{H}_R, \hat{X}_A] 
 = \epsilon_A\frac{\hat{P}_R}{M} = \epsilon_A \partial_t \hat{X}_R\,,
 \\
 \delta_{\epsilon_A} \hat{P}_A &= i\epsilon_A [ \hat{H}_R, \hat{P}_A] 
 = -\epsilon_A V'(\hat{x}_R)\,.
\end{align}
Here note that the $\hat{X}_A$ variable is mapped to 
the time derivative of $\hat{X}_R$ in particular.
From the MSR point of view,  this is reflected to the fact that 
the MSR effective action \eqref{eq:MSR2} is \textit{not} invariant 
under the infinitesimal transformation labelled by $\epsilon_A$:
\begin{equation}
 \begin{cases}
  X_R (t) \to X_R' (t) = X_R (t), \\ 
  X_A (t) \to X_A' (t) = X_A (t) + \epsilon_A \dot{X}_R .
 \end{cases}
 \label{eq:Timetr2}
\end{equation}
Since we have performed the integration over $P_R$ and $P_A$ to get the 
action in the configuration space, we only have the transformation 
with respect to $X_R$ and $X_A$.
The variation 
$\delta_{\epsilon_A} S_{\msr} 
\equiv S_{\msr} [X_R', X_A'] - S_{\msr} [X_R, X_A]$ 
under Eq.~\eqref{eq:Timetr2} then reads
\begin{equation}
 \begin{split}
  \delta_{\epsilon_A} S_{\msr} 
  &= - \epsilon_A \int dt
  \Big[ i M \dot{X}_R \ddot{X}_R + i\dot{X}_R V'(X_R) 
  + M \gamma \dot{X}_R^2 + A X_A \dot{X}_R \Big]
  + O \big( \epsilon_A^2 \big) \\
  &= - i \epsilon_A \int dt
  \frac{d H_R}{dt} 
  -  \epsilon_A \int dt
  \Big[ M \gamma \dot{X}_R^2 + A X_A \dot{X}_R \Big]
  + O \big( \epsilon_A^2 \big) .
 \end{split}
\end{equation}
Although the first term is the total derivative of the Hamiltonian 
which respects symmetry under \eqref{eq:Timetr2}, 
the second term does not vanish in the presence of fluctuation and dissipation. 

\medskip
We can summarize the above result as follows. 
When we consider open systems like the Brownian particle system, 
we have two time-translational symmetries 
which we call the $H_R$-symmetry and the $H_A$-symmetry: 
the former is explicitly broken and the latter is not. 
The conserved charges, or generators associated with them, are 
the original Hamiltonian and the Fokker-Planck Hamiltonian, respectively.
This doubled symmetry structure is naturally understood from underlying 
quantum theory with the help of the Schwinger-Keldysh formalism. 
We will see this in Sec.~\ref{micro:origin}.

\medskip 
In the end of this subsection, let us discuss spontaneous symmetry breaking
(SSB) of the doubled time-translational symmetry. 
As we elaborated, one symmetry ($H_R$ symmetry) is explicitly broken 
since the system Hamiltonian is no longer conserved 
due to the nonvanishing fluctuation and dissipation, and thus, 
we focus on the other symmetry ($H_A$-symmetry) generated by 
the Fokker-Planck Hamiltonian $\hat{H}_A$.
This symmetry structure is a striking feature of the open systems where 
we can break the conservation law in the presence of the environment\footnote{
Nevertheless, we note that there is a situation where we do have 
conservation laws 
even though we have the fluctuation and dissipation.
One such example is the EFT for dissipative hydrodynamics (See e.g. Refs.~\cite{Crossley:2015evo,Glorioso:2017fpd} for a detailed discussion).   
Although we have the conservation law 
in this kind of the closed dissipative systems, 
the corresponding symmetry is broken due to the close-time-path nature of 
the Schwinger-Keldysh formalism. 
As a consequence, that symmetry (e.g. $H_R$-symmetry) is nonlinearly 
realized and the low-energy effective theory is described by a kind 
of the NG mode. }.
It is worthwhile to point out that 
when the fluctuation and dissipation in our open systems are small, 
the $H_R$-symmetry can be thought of as an approximate symmetry, 
and we may construct the effective field theory for such weakly open systems 
with the help of the spurion field.
In this paper, we only focus on the strongly open system 
which explicitly breaks the energy conservation ($H_R$-symmetry), 
and consider spontaneously broken symmetry of the $H_A$-symmetry 
in the following sense. 

\medskip
Recalling the Sch\"odinger-like equation \eqref{eq:sch}, 
we put a \textit{weak} criterion%
\footnote{
Here we call this criterion \textit{weak} to distinguish a usual 
(\textit{strong}) definition of SSB for the vacuum or ground state of systems.
In the usual situation, we act the Noether charge, 
or more precisely, the commutation relation between the Noether charge and 
order parameter field to the vacuum state to judge whether SSB occurs or not. 
Since we do not consider the vacuum state here, we call our criterion 
as \textit{weak}.
\label{footnote4}}
for SSB by operating the Noether charge $\hat{Q}$---the Fokker-Planck 
Hamiltonian $\hat{H}_A$ in our situation---on the probability distribution 
$\Pcal$: if $\hat Q \Pcal = 0$, our system remains symmetric 
while $\hat Q \Pcal \neq 0$ implies that SSB takes place.
For a stationary distribution $\Pcal_{\rm eq}$, we have 
\begin{align}
 \hat H_A \Pcal_{\rm eq} = 0.\label{nonssbeq}
\end{align}
This implies that a stationary distribution is literally 
time-translationally invariant, 
and hence, if the minimum eigenvalue of $\hat{H}_A$ is zero, our system falls into the time-translationally symmetric phase after a while.
On the other hand, 
for a general time dependent probability
distribution $\Pcal$, one finds
\begin{align}
 \hat H_A \Pcal \neq 0. \label{ssbnoneq}
\end{align}
Then, from our (weak) criterion, 
we can interpret the time-dependent probability distribution as 
a symmetry broken phase.
Regardless of our \textit{weak} definition of SSB,
Eq.~(\ref{ssbnoneq}) plays an essential role since it provides 
a starting point to construct the EFT
for time-translational symmetry breaking in open systems;
Eq.~(\ref{ssbnoneq}) together with e.g. a slow-roll approximation
says that low-energy dynamics of non-stationary systems can be described by 
effective field theory of the Nambu-Goldstone field\footnote{
Strictly speaking, it might be not appropriate to call it the Nambu-Goldstone field, because we are employing the weak criterion for SSB. However, this field transforms nonlinearly under the spontaneously broken symmetry, hence it follows the same property as the ordinary Nambu-Goldstone field from the representation theory perspective. We therefore call it the Nambu-Goldstone field in this paper.},
which represents a perturbation around
the time-dependent background.

\subsubsection{Dynamical KMS symmetry}
\label{subsec:KMS_B}

In this subsection, we comment on 
a way to implement the fluctuation-dissipation relation (FDR) 
to the configuration-space MSR effective action 
by the use of the discrete symmetry.
As we mentioned just after Eq.~\eqref{eq:NoiseAvg}, the FDR $A= 2M \gamma T$ would hold for the systems 
that obey a detailed balance condition.
Furthermore, if our system initially (at $ t = - \infty$) 
stays in a thermal equilibrium state, 
we take average over the equilibrium probability distribution 
$\mathcal P_{\mathrm{eq}} \propto e^{- \beta H_R (t = - \infty)}$ with an inverse 
temperature $\beta = 1/T$. 
Then, the total MSR effective action leads to
\begin{align}
\nonumber
 &iS_{\rm MSR} [X_{R},X_{A}]  - \beta H_R (t=-\infty)
 \\
& = \int_{-\infty}^{\infty} dt\, 
 \Big[  -i X_A \big( M\ddot{X}_R + M \gamma\dot{X}_{R} + V'(X_{R})
 \big) - M \gamma T X_A^2 \Big] 
 - \beta H_R (t=-\infty), \label{eq:MSR+H}
\end{align}
where we explicitly wrote the integral region. 
In this case, there is a nontrivial relation between the low-energy 
(Wilson) coefficients from the EFT viewpoint. 
As a consequence of this,
we have an additional discrete symmetry 
known as the dynamical KMS (Kubo-Martin-Schwinger) symmetry~\cite{Aron,Glorioso:2017fpd,Aron2,Sieberer:2015hba}
generated by
\begin{equation}
 \begin{cases}
  X_R (t) \to X'_R (t) = X_R(-t) , \\
  X_A (t) \to X'_A (t) = X_A (-t) + i \beta \partial_{-t} X_R (-t).
 \end{cases}\label{dyn:KMS:x}
\end{equation}
Here, instead of $\dot{X}_R$ we use $\partial_{-t}$ to avoid confusion.
Note that we can think of the dynamical KMS transformation as the sequence of the $H_R$ transformation with $\epsilon_{A}=-i\beta$ and time-reversal 
transformation. 
This point of view is useful to consider the dynamical KMS 
transformation for the NG fields in the next section.
We note that if we again act the dynamical KMS transformation, 
it becomes identity transformation, and thus, 
the dynamical KMS symmetry is $\bm{Z}_2$ symmetry.
We can directly show invariance of the MSR action \eqref{eq:MSR+H} 
under this transformation as follows:
\begin{equation}
 \begin{split}
  &\quad iS_{\textrm{MSR}} [X'_R,X'_A] - \beta H_R \big( X_R'(t=-\infty) \big) 
  \\
  &= \int_{-\infty}^{\infty} dt\, 
  \Big[ -i \big( X_A (-t) + i \beta \partial_{-t} X_R (-t) \big) 
  \big( M \partial_t^2 X_R (-t) + M\gamma \partial_t X_{R} (-t)+ V'(X_{R}(-t))
  \big)  \\
  &\hspace{55pt} 
  - M \gamma T \big( X_A (-t) + i \beta \partial_{-t} X_R (-t) \big)^2  \Big] 
  - \beta H_R \big( X_R(t=\infty) \big)
  \\
  &= \int_{-\infty}^{\infty} dt\, 
  \Big[  -i  X_A  
  \big( M \partial_t^2 X_R  + M \gamma \partial_t X_{R} + V'(X_{R}) \big)  
  - M \gamma T  X_A^2  
  + \beta \partial_t H_R \Big] 
  - \beta H_R ( t=\infty )
  \\
  &= iS_{\mathrm{MSR}} [X_R,X_A] - \beta H_R (t= -\infty).
 \end{split} 
\end{equation}
This additional discrete symmetry emerges once we assume the FDR. 
In other words, if we demand the dynamical KMS symmetry, it brings about 
the nontrivial relation between some low-energy coefficients 
of terms in the effective action, 
which may be a manifestation of the generalized fluctuation-dissipation theorem.

\subsection{Microscopic origin of symmetries}
\label{micro:origin}

From the microscopic point of view, the fluctuation and dissipation 
in the Brownian motion
originate from collisions with molecules of the surrounding fluid. 
The Langevin equation can then be thought of as an effective equation of motion 
for the Brownian particle taking into account such a coupling 
with an environment. 
More generally, the fluctuation and dissipation in dissipative systems 
 arise as a consequence of coarse-graining of 
the microscopic dynamics. 
The Schwinger-Keldysh formalism provides a nice framework 
to capture such an origin of fluctuation and dissipation 
from underlying quantum theory. 
We here elaborate the microscopic origin of the two 
(broken and unbroken) symmetries mentioned 
in the previous subsections based on the Schwinger-Keldysh formalism.

\begin{figure}[t]
\begin{center}
\includegraphics[width=8cm]{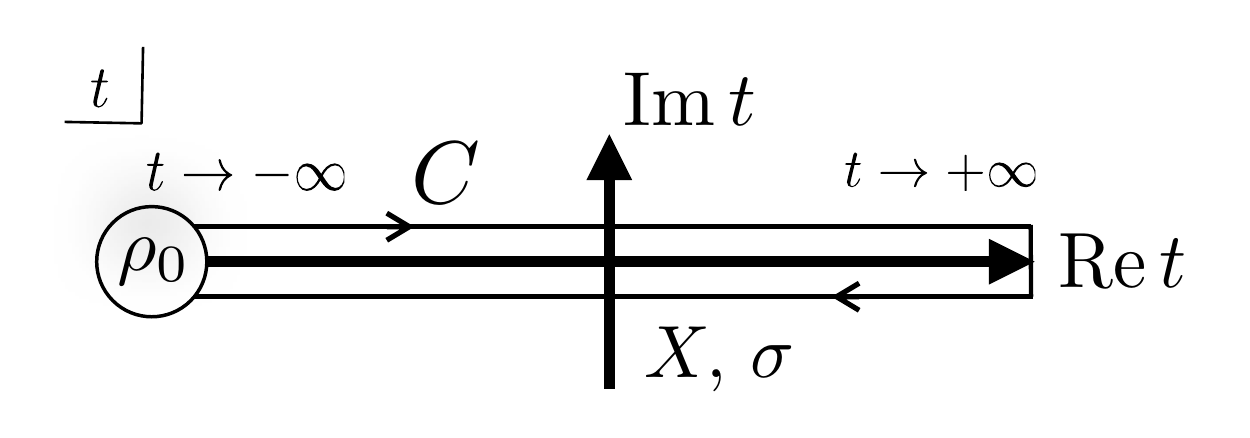}
\caption{
 The closed-time-path on which the generating functional is defined.
 When we assume that the initial density operator $\hat{\rho}_0$ is 
 given by a thermal density operator: $\hat{\rho}_0 \propto e^{-\beta \hat{H}}$,
 we have another path to the negative imaginary time (See Appendix~\ref{kim}).}
 
\label{fig:CTP}
\end{center}
\end{figure}

\medskip
Suppose that we know the microscopic theory behind the Brownian motion 
and have a Hamiltonian description of the full quantum system 
including both the Brownian particle and the surrounding 
light particles---components of a fluid the fluctuation and 
dissipation originating from. 
For the full description of this system, the closed-time-path (CTP) 
generating functional with an initial density operator $\hat{\rho}_0$
is defined by
\begin{align}
 Z[J_1, J_2] 
 = 
 {\rm Tr} \Big[\hat{\rho}_0 \,
 \hat{U}_{J_2}^\dagger(\infty,-\infty)
 \hat{U}_{J_1} (\infty,-\infty) \Big]\,, 
 \label{partition}
\end{align}
where unitary operators $\hat{U}_{J_{a}}(t_2, t_1)$ denote 
time evolution operators 
from $t_1$ to $t_2$ in presence of the external source $J_a(t)~(a=1,2)$.
Note that $\hat{U}_{J=0} $ is generated by the Hamiltonian of the total system, and $\hat{U}_J $ is also unitary even in the presence of external fields as long as the source term is Hermitian. 
Here, introducing two types of sources $J_a~(a=1,2)$ enables us to calculate 
all types of correlation functions---retarded, advanced, and symmetric Green functions---necessary to describe real-time dynamics.
Otherwise, the CTP generating functional becomes trivial due to
the unitarity of the time evolution operator:
$\hat{U}_{J} (t_2,t_1) U_{J}^\dagger(t_2,t_1) = 1$. 
The corresponding path-integral formula is given by
\begin{align}
 Z[J_1,J_2]
 =\int \mathcal{D} X_1 \mathcal{D} X_2
 \mathcal{D} \sigma_1 \mathcal{D} \sigma_2
 \exp 
 \Big[iS_{\rm micro}[X_1,\sigma_1;J_1]-iS_{\rm micro}[X_2,\sigma_2;J_2] \Big]  
 \rho_0 (X,\sigma) \label{pi_micro}  \,,
\end{align}
where $S_{\rm micro}[X,\sigma;J]$ is the microscopic action 
for the position of the Brownian particle $X$ 
and other microscopic degrees of 
freedom for the environment collectively denoted by $\sigma$ 
with the external source $J$. 
Here $\rho_0 (X,\sigma)$ denotes an initial ensemble 
determined by $\hat{\rho}_0$.
Note that the mixing between the variables with different indices ($a=1,2$)
is suppressed by an infinitesimal parameter $\epsilon$ of the $i\epsilon$ 
prescription (See the discussion in Appendix~\ref{kim}).
For later convenience, let us only turn on the external source 
for the Brownian particle $X$ and turn off that for the environment $\sigma$. 
Just as we introduced a pair of sources $J_a$, 
the dynamical variables $X$ and $\sigma$ are doubled, so that we have 
 $X_a$ and $\sigma_a~(a=1,2)$. 
As depicted in Fig.~\ref{fig:CTP}, the $a=1$ ($a=2$) variables are responsible for the time-evolution from the past (future) to the future (past).

\medskip
When the microscopic action $S_{\rm micro}$ enjoys time-translational symmetry, the exponent in Eq.~\eqref{pi_micro} is invariant under 
doubled time-translations 
labeled by $\epsilon_a~(a=1,2)$\footnote{
To be precise, the transformation $\epsilon_1\neq\epsilon_2$ is violated by the $\mathcal{O}(\epsilon)$ mixing between the $1$ and $2$ variables mentioned earlier, where $\epsilon$ without subscripts $a=1,2$
is an infinitesimal parameter for the $i\epsilon$ prescription.},
\begin{align}
 X_1(t) \to X'_1 (t) = X_1 (t+\epsilon_1)
 \,,
 \quad
 X_2(t)\to X'_2 (t) = X_2 (t+\epsilon_2)
\end{align}
with a similar transformation rule for $\sigma_a$ and $J_a$, 
because the $a=1$ variables and $a=2$ variables are separated. 
As we elaborate below, combinations of 
these two symmetries turn out to be the two symmetries 
associated to $\hat{H}_{R}$ and $\hat{H}_A$ mentioned 
in the previous subsection. 
We then discuss how such a symmetry structure changes after integrating out 
the environment $\sigma$. 
The crucial point here is that $\sigma_1$ and $\sigma_2$ has 
a nontrivial correlation related to on-shell particle creations. 
As a result, the effective action after integrating out $\sigma_a$ 
cannot be separated into the $a=1$ and $a=2$ parts if 
the Brownian particle $X_a$ is coupled to the environment:
\begin{align}
 Z[J_1,J_2]=\int \mathcal{D} X_1 \mathcal{D} X_2 
 \exp \Big[iS_{\rm eff}[X_1,X_2;J_1,J_2] \Big]
 \,.\label{eq:MSR10}
\end{align}

\begin{figure}[t]
 \begin{center}
  \includegraphics[width=15cm]{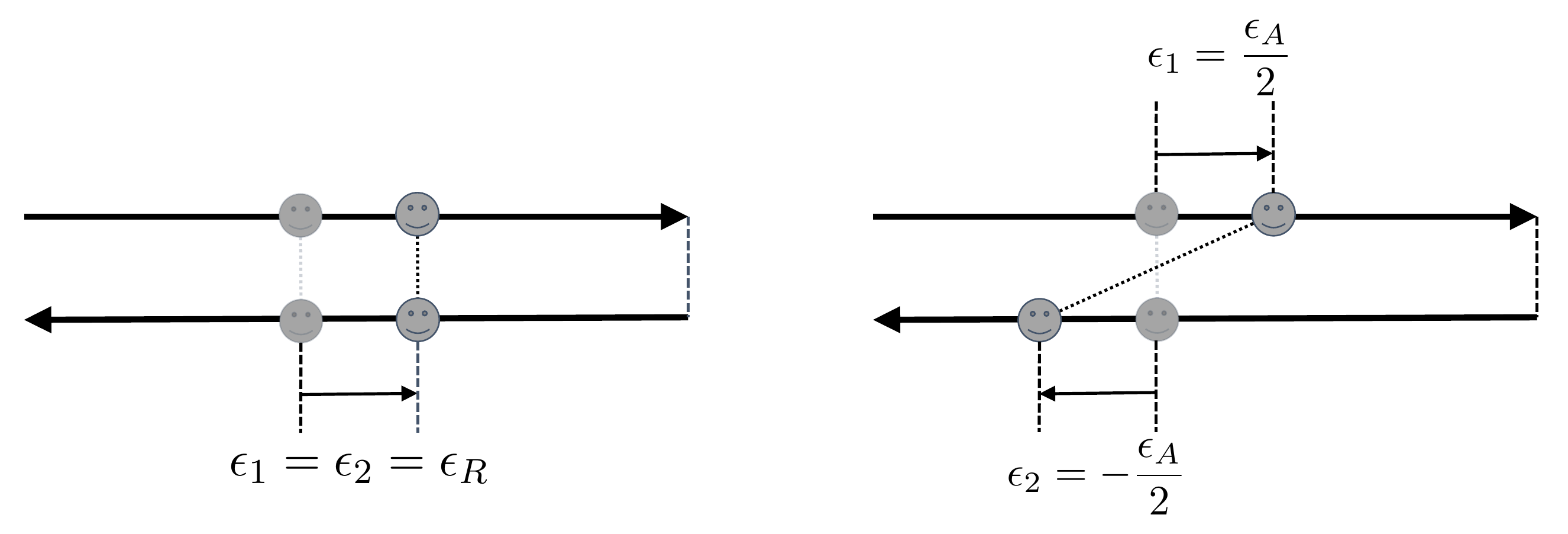}
  \caption{
  $\epsilon_{R}$ and $\epsilon_{A}$ transformations on the closed time path.
  $\epsilon_{R}$ translates the $a=1,2$ variables in the same direction 
  while $\epsilon_{A}$ does in the opposite directions.
  Lagrangians without explicit $t$ dependence is invariant 
  under $\epsilon_{R}$ translations only.
As depicted in the right panel, equal-time 1-2 couplings are not equal-time anymore after the $\epsilon_A$-transformation.}
  \label{fig:TwoTimeTr}
 \end{center}
\end{figure}

Let us see a famous concrete example and explicitly 
obtain $S_{\mathrm{eff}}[X_1,X_2]$ for the Brownian particle. 
We consider a one dimensional Browninan point particle $X(t)$ 
that couples to the light harmonic oscillator 
$\sigma (t) =\{x_1 (t), \cdots, x_N(t)\}$, whose total action is given by
\begin{equation}
 S_{\mathrm{total}}[X,\{x_n\}] 
  = \int dt 
  \Big[ L_{\mathrm{sys}} + L_{\mathrm{env}} + L_{\mathrm{int}} \Big] .
  \label{eq:TotalL}
\end{equation}
Here $L_{\mathrm{sys}},\, L_{\mathrm{env}}$, and $L_{\mathrm{int}}$ 
respectively denote Lagrangians for the Brownian particle, 
environment harmonic oscillators and their interactions:
\begin{equation}
 L_{\mathrm{sys}} 
  = \dfrac{M}{2} \dot{X}^2 - V (X), 
  \quad
  L_{\mathrm{env}} 
  =  \displaystyle{\sum_{n=1}^N }
  \left[ \dfrac{m_n}{2} \dot{x}_n^2 - \dfrac{1}{2} m_n \omega_n^2 x_n^2 \right], 
  \quad
  L_{\mathrm{int}} 
  =  \displaystyle{\sum_{n=1}^N} g_nx_n X , 
  \label{eq:EachL}
\end{equation}
with the mass $m_n$ and the frequency $\omega_n$ of the $n$-th 
environment harmonic oscillator.
For simplicity, we introduced linear couplings between $X$ and $x_n$ with the coupling constants $g_n$.
We employ the so-called \textit{Ohmic bath}~\cite{kamenev2011field} as the environment, which accommodates a continuous spectrum, hence $N$ is very large (see Appendix~\ref{kim} for details).
We also assume thermal equilibrium at initial time ($t=-\infty$).
In Appendix~\ref{kim}, starting from this Lagrangian, we derive the following effective action 
for the heavy particle
\begin{align}
 iS_{\mathrm{eff}} [X_1,X_2] 
 = i\int dt\, 
 &\left[ - \frac{M}{2} X_{i} D^{-1}_{ij} X_{j} - V(X_1) + V(X_2) \right], 
 \label{eq:MSR3}
\end{align}
where we defined 
\begin{align}
 D^{-1}_{ij}
 = \begin{pmatrix}
    \partial_{t}^{2}-\dfrac{iA}{M} & \gamma\partial_{t} + \dfrac{iA}{M} 
    \vspace{5pt} \\
    - \gamma\partial_{t} + \dfrac{iA}{M} & - \partial_{t}^{2} - \dfrac{iA}{M}
   \end{pmatrix}
   \quad
   {\rm with}
   \quad
   A=2M\gamma/\beta
   .
  \label{D-1}
\end{align}
Here $\gamma$ ($>0$) is a parameter characterizing the frequency distribution of the harmonic oscillators in the Ohmic bath. Note that the FDR $A=2M\gamma/\beta$ is satisfied because we assumed thermal equilibrium.
As we expected, we have mixing terms after integrating out 
the environment $x_n$, and thus, $a=1,2$ variables couple with each other.
Because of this mixing term, 
the effective action is not invariant 
under the choice of $\epsilon_1 \neq \epsilon_2$ anymore. 
The symmetry of the effective action is then the time-translational symmetry 
with a restricted parameter 
(See the left picture in Fig.~\ref{fig:TwoTimeTr})
\begin{align}
\label{eR}
 \epsilon_1 = \epsilon_2 = \epsilon_R 
 \,,
\end{align}
where we labeled the same parameter as $\epsilon_R$. 
For later use, we also label the (explicitly) broken symmetry by $\epsilon_A$ as
(See the right picture in Fig.~\ref{fig:TwoTimeTr})
\begin{align}
\label{eA}
 \epsilon_1 = - \epsilon_2 = \frac{\epsilon_A}{2} \,.
\end{align}
In this way, the two time-translational symmetries of the microscopic action is broken into the diagonal one $\epsilon_1=\epsilon_2$ 
in the presence of the fluctuation and dissipation.
This is precisely what we encountered in the previous subsection.

\medskip
To see the relation to the argument in the previous subsection more explicitly,
it is convenient to introduce the following new basis:
\begin{align}
 X_R = \frac{X_1 + X_2}{2}\,,
 \quad
 X_A= X_1 - X_2  \,.
\end{align}
Under the transformation~\eqref{eR} parameterized by $\epsilon_R$, they transform as
\begin{align}
 \begin{cases}
  X_R(t) \to X'_R (t) = X_R (t+\epsilon_R)\,, \\
  X_A(t) \to X'_A(t) = X_A (t+\epsilon_R)\,,
 \end{cases}
\end{align}
which is nothing but the $H_A$ transformation \eqref{eq:Timetr1} generated by 
the Fokker-Planck Hamiltonian, under the identification of 
$X_{R}$ and $X_A$ with those in the MSR formalism. 
On the other hand, the explicitly broken one~\eqref{eA}, parameterized by $\epsilon_A$, 
takes the form,
\begin{equation}
  \begin{cases}
   X_R(t) 
   \to X'_{R}(t) 
   = X_R(t)+ \dfrac{1}{4} \epsilon_A \dot{X}_A(t) + O (\epsilon_A^2) \,,
   \vspace{5pt} \\
   X_A(t) 
   \to X'_A(t) 
   = X_A(t) +\epsilon_A\dot{X}_R+ O (\epsilon_A^2)\,.
  \end{cases}
\end{equation}
Here it is convenient to assign $\hbar$-dependence such that 
$X_{R} =O (\hbar^0)$ and $X_A = O(\hbar)$. 
Then, in the semiclassical limit ($\hbar\to0$), 
the above is reduced to the following form,
\begin{equation}
 \begin{cases}
  X_R(t) \to 
  X'_R(t) = X_R(t) + O(\epsilon_A^2)\,,
  \\
  X_A(t)\to 
  X'_A(t) = X_A(t) + \epsilon_A \dot{X}_{R} + O(\epsilon_A^2)\,,
 \end{cases}
\end{equation}
the leading-order of which 
is nothing but the $H_R$ transformation~\eqref{eq:Timetr2}
generated by the original Hamiltonian of the Brownian particle.
This implies that the MSR effective action can be obtained 
as the semiclassical limit of the Schwinger-Keldysh action.
\medskip

To summarize, the microscopic (UV) action 
in the Schwinger-Keldysh formalism enjoys doubled 
time-translational symmetries labeled by 
$\epsilon_R$ and $\epsilon_A$, just like all the fields are doubled 
because of the closed-time-path nature. 
Nevertheless, 
the effective (IR) action after integrating out the environment accommodates 
mixings between $a=1,2$ variables, which are interpreted as 
stochastic and dissipative effects. 
Such mixings explicitly break the doubled symmetries into the diagonal one 
labeled by $\epsilon_R$, which is nothing but the $H_A$ time-translational 
transformations generated by the Fokker-Planck Hamiltonian. 
On the other hand, the explicitly broken symmetry labeled by $\epsilon_A$ 
is identified with the $H_R$ transformation generated by the Hamiltonian of 
the Brownian particle in the semiclassical limit ($\hbar \to 0$). 
Then, we are now able to discuss the potential to describe the non-stationary 
probability distribution in a view of effective field theory 
based on the spontaneous symmetry breaking of the 
$H_A$-symmetry, 
which is the main content discussed in the next section.

\section{Constructing EFT based on Schwinger-Keldysh formalism}
\label{section:ConstructingEFT}

So far we have elucidated the symmetry structure for the one dimensional 
Brownian particle, and shown that it can be naturally understood from the 
Schwinger-Keldysh viewpoint as summarized in Table~\ref{table:ss}.
In this section, promoted by the previous  discussion---in particular, 
our \textit{weak} criterion of SSB labelled by $\epsilon_R$---we regard the 
non-equilibrium situations with a time-dependent condensate as a 
broken phase of time-translational symmetry.
In Sec.~\ref{sec:Prelim}, 
we first review the effective Lagrangian for time-translational symmetry 
breaking in the in-out formalism and the basic recipe for EFT 
based on the Schwinger-Keldysh (in-in) formalism. 
In Sec.~\ref{subsec:EFT_const}, we construct the general effective Lagrangian
for Nambu-Goldstone fields based on the symmetry structure. 
We derive the dispersion relation for the NG bosons and 
also consider the restriction for EFT from the dynamical KMS symmetry. 
Furthermore, in Sec.~\ref{sec:ModelAnalysis}, 
we consider the simple UV model composed of a single scalar field 
with noises and dissipations and perform the tree-level analysis.
This model analysis illustrate how low-energy coefficients in the 
constructed EFT are related to information on UV theory.
Throughout this section, we assume that our system is originally 
Lorentz invariant before symmetry breaking, unless otherwise stated.

\begin{table}[t]
  \centering
  \begin{tabular}{c|cc}
    \hline
    &$H_{A}$-symmetry&$H_{R}$-symmetry\\
    \hline \hline
   Open system (stationary) & Unbroken & Explicitly broken \\
 Open system (non-stationary) & Spontaneously broken & Explicitly broken  \\
    \hline
  \end{tabular}
 \caption{Symmetry structure in open systems: Because of the doubled symmetry in the Schwinger-Keldysh formalism, we need to discuss the $H_A$ and $H_R$ symmetries separately. In open systems, the $H_R$ symmetry is explicitly broken, whereas the $H_A$ symmetry is unbroken/spontaneously broken in the stationary/non-stationary backgrounds.
 It is in a sharp contrast to the closed system at the zero temperature, where the dynamics is described by the Hamiltonian of the total system.
 }
 \label{table:ss}
\end{table}

\subsection{Preliminaries}
\label{sec:Prelim}
In Sec.~\ref{subsec:EFT_in-out}, 
we first summarize a basic construction of the effective field theory 
for time-translational symmetry breaking, which has been developed 
in the context of cosmic inflation~\cite{Cheung:2007st}.
We then present a way to extend the EFT to open systems 
in non-equilibrium situations based on the Schwinger-Keldysh formalism.

\subsubsection{Effective Lagrangian in the in-out formalism}
\label{subsec:EFT_in-out}

Let us begin with a brief review on the in-out effective Lagrangian 
for time-translational symmetry breaking. 
Considering \textit{any} UV theory with a time-translationally invariant action $S_{\mathrm{micro}}$,
we suppose that a certain scalar condensate 
$\langle\phi(t,\mathbf{x})\rangle$ has a nontrivial time-dependence:
\begin{align}
\label{condensate}
 \langle\phi(t,\mathbf{x})\rangle
 = \bar{\phi}(t)
\quad
{\rm with}
\quad
\dot{\bar{\phi}}\neq0
\,.
\end{align}
In this situation, our system clearly breaks time-translational symmetry,
and thus, 
the NG mode $\pi(x)$ can be embedded into this time-dependent condensate as
\begin{align}
\label{embedding}
\phi(t,\mathbf{x})=\bar{\phi} \big(t+\pi(t,\mathbf{x}) \big)
\end{align}
with the following transformation rule under translations:
\begin{align}
 \label{translations}
 \pi(t,\mathbf{x})
 \to \pi'(t,\mathbf{x})
 = \pi(t+\epsilon^0,\mathbf{x}+\boldsymbol{\epsilon})+\epsilon^0\,.
\end{align}
In addition, under the Lorentz transformation, the NG mode $\pi(x)$ transforms as
\begin{align}
 \label{Lorentz}
 \pi(t,\mathbf{x})
 \to \pi' (t,\mathbf{x}) 
 = \pi \left(\Lambda^0{}_\mu x^\mu,\Lambda^i{}_\mu x^\mu\right)+\Lambda^0{}_\mu x^\mu-t
\,,
\end{align}
where we introduced $\Lambda^\mu{}_\nu\in SO(1,3)$.
Note that, as is usual for the NG mode,
$\pi(x)$ nonlinearly transforms under time-translations and boosts, 
which are broken by the time-dependent condensate~\eqref{condensate}\footnote{
As is known as the inverse Higgs effect~\cite{Ivanov:1975zq} 
(See also, e.g.,~\cite{Nicolis:2013sga,Endlich:2013vfa,Brauner:2014aha,Hidaka:2014fra} for recent discussion), 
the number of massless NG bosons does not necessarily coincide with 
that of broken spacetime symmetries. As a result, 
the broken boost symmetry can be nonlinearly realized 
without introducing NG bosons for boosts.}.
Even though we considered a single scalar condensate as an illustrative example,
the above transformation rule is applicable in general.

\medskip
As the embedding~\eqref{embedding} suggests, 
the general effective action is constructed from~\cite{Cheung:2007st}
\begin{align}
 t + \pi(x)
\quad
\text{and its derivatives.}
\end{align}
Indeed, we can explicitly show that these ingredients are invariant 
under the transformations~\eqref{translations}-\eqref{Lorentz} upon 
an appropriate coordinate transformation. 
For the construction of effective Lagrangian, it is convenient to introduce
\begin{align}
P_\mu=\partial_\mu(t+\pi)=\delta^0_\mu+\partial_\mu\pi\,,
\label{notation:P}
\end{align}
which is used as a basic building block.
In particular, its Lorentz invariant square is then
\begin{align}
 P_\mu P^\mu=-1-2\dot{\pi}+(\partial_\mu\pi)^2\,.
\end{align}
At the leading order in derivative expansion\footnote{
We dropped higher derivative terms containing $\pi$ with two or 
more derivatives.},
the general effective action is given by~\cite{Cheung:2007st}
\begin{align}
\label{S_in-out_P}
 S_{\text{in-out}} 
 = -\frac{1}{2}\int d^4x 
 \left[ \alpha_0(t+\pi) + \alpha_1(t+\pi) P_\mu P^\mu 
 + \sum_{n\geq2}\alpha_n(t+\pi)(P_\mu P^\mu+1)^n 
\right]\,,
\end{align}
where $\alpha_n$'s are arbitrary functions of $t+\pi$. In contrast to the internal symmetry breaking case, 
the NG boson $\pi$ does not enjoy the shift symmetry in general. 
As a result, the above action contains terms linear in $\pi$:
\begin{align}
 S_{\text{in-out}} 
 = -\frac{1}{2}\int d^4x 
 \left[ \alpha_0(t) - \alpha_1(t)
 + \Big(\alpha'_0(t)-\alpha'_1(t)\Big)\pi-2\alpha_1(t)\dot{\pi} + O (\pi^2)
\right]\,.
\end{align}
To remove such a tadpole term, we use the background equation of motion to require $\alpha'_0=-\alpha'_1$. The action can then be reformulated as
\begin{align}
S_{\text{in-out}}=-\frac{1}{2}\int d^4x\left[\alpha_1(t+\pi)(\partial_\mu\pi)^2+\sum_{n\geq2}\alpha_n(t+\pi)\left(-2\dot{\pi}+(\partial_\mu\pi)^2\right)^n
\right]\,,\label{action:pi:inout}
\end{align}
where we chose the integration constant of the equation $\alpha'_0=-\alpha'_1$ 
such that $\alpha_0=-\alpha_1$. This is the general in-out effective action 
of the NG boson for the broken time-translation.

\medskip
Here it is worth emphasizing that the EFT coefficients $\alpha_n$'s 
in Eq.~\eqref{action:pi:inout} are generally 
time-dependent in contrast to the internal symmetry breaking case.
The EFT is therefore less predictive at this stage because of the functional degrees of freedom. Also, the energy is not well-defined because there is no time-translational invariance anymore.
In order to make the energy well-defined and provide the universal
predictive power to the EFT, it is convenient to impose another symmetry assumption. One typical example is to neglect the time-dependence of the order parameter\footnote{
Another interesting situation is when the order parameter is periodic in time like the 
synchronization phenomena or the time crystal.
In such a case, energy is well defined in the long time range. 
The NG field will accommodates a band structure, 
just as the periodic modulation along spatial direction~\cite{Hidaka:2015xza}. We will revisit this issue elsewhere.}.
In analogy with the cosmic inflation\footnote{During the slow-roll inflation, the velocity of the inflaton 
background $\dot{\bar{\phi}}$ is nearly constant. 
As a result, there exists an approximate de Sitter time-translational symmetry.},
we call this parameter regime the slow-roll regime.

\medskip
In the slow-roll regime, we may neglect 
the time-dependence of $\alpha_n$'s, and the effective action for the NG mode 
is simplified into the following form:
\begin{align}
 \label{slow-roll_in-out}
 S_{\text{in-out}} 
 = - \frac{1}{2} \int d^4x
 \left[ \alpha_1 (\partial_\mu\pi)^2 + 
 \sum_{n\geq2} \alpha_n \left(-2\dot{\pi} + (\partial_\mu\pi)^2 \right)^n
 \right]\,,
\end{align}
where $\alpha_n$'s are now constant parameters
and $\alpha_1$ characterizes the symmetry breaking scale.
Note that the first term is just a canonical kinetic term multiplied 
by an overall factor $\alpha_1$, so that the NG mode enjoys a relativistic dispersion relation and has no self-interactions unless the higher-order terms $\alpha_n$ are turned on. In general, the NG mode has a linear dispersion with a non-unity propagating speed as
\begin{equation}
 \omega^2 = c_s^2 k^2 \with 
  c_s^2 \equiv \frac{\alpha_1}{ \alpha_1 - 4 \alpha_2}\,.
\end{equation}
We will see how this linear dispersion relation is modified when 
we take into account open system effects such as fluctuation and dissipation. 

\medskip
Before closing this subsection, we comment on a transformation property 
of the NG mode in the slow-roll regime.
As we can see from Eq.~\eqref{slow-roll_in-out}, 
the NG boson enjoys a shift symmetry in the slow-roll regime.  
This can be also rephrased as that the slow-roll regime is realized 
by assigning a {\it linear} transformation rule for $\pi(x)$ 
under the translations 
\begin{align}
 \label{linear_translations}
 \pi(t,\mathbf{x}) 
 \to \pi'(t,\mathbf{x}) 
 = \pi (t + \epsilon^0,\mathbf{x} + \mathbf{\epsilon}) \,,
\end{align}
instead of the nonlinear one~\eqref{translations}. 
Since the linear transformation rule~\eqref{linear_translations} 
is the same as the one for ordinary matter scalars, 
one may wonder that we can also add $\pi^{n}$ 
to the effective Lagrangian. 
Nevertheless, it is indeed not true. 
This is because we still assign the nonlinear 
boost transformation rule~\eqref{Lorentz}\footnote{
The NG mode in the relativistic superfluid also enjoys 
the transformation rule, \eqref{linear_translations} and \eqref{Lorentz}, 
so that its effective action is given by Eq.~\eqref{slow-roll_in-out} 
in the relativistic setup.}.
As a result, the NG mode has to be always accompanied by derivatives,
and then ingredients of the effective action 
turn out to be derivatives of $t+\pi(x)$.
In other words, $\partial_{\mu}\pi$ is not covariant under the nonlinear boost transformation.
At the leading order in derivative expansion, 
the effective action is thus given by~\eqref{slow-roll_in-out}.

\subsubsection{Recipe for Schwinger-Keldysh-based EFT}
\label{subsec:EFT_in-in}

As is already introduced in section~\ref{micro:origin}, 
the most important quantity in the Schwinger-Keldysh formalism is 
the closed-time-path generating functional (CTPGF) $Z[J_1, J_2]$. 
While the CTPGF contains all information on real-time dynamics of our theory, 
we are only interested in the low-energy dynamics associated with 
time-translational symmetry breaking. 
Therefore, integrating out all UV degrees of freedom except for the 
doubled NG-like mode $\pi_a(x) ~(a=1,2)$, 
we would like to construct the EFT for them as follows:
\begin{align}
\nonumber
  Z[J_1, J_2] 
  &= \mathrm{Tr} \Big[\hat{\rho}_0 \,
  \hat{U}_{J_2}^\dagger(\infty,-\infty)
  \hat{U}_{J_1} (\infty,-\infty) \Big]\,  \\
\nonumber
  &=\int \mathcal{D} \varphi_1 \mathcal{D} \varphi_2
  \mathcal{D} \sigma_1 \mathcal{D} \sigma_2
  \exp \Big[
  iS_{\mathrm{micro}} [\varphi_1,\sigma_1;J_1] 
 - iS_{\mathrm{micro}} [\varphi_2,\sigma_2;J_2] \Big]  
  \rho_0 (\varphi,\sigma) 
  \, \\
  &= \int \mathcal{D} \pi_1 \mathcal{D} \pi_2
  \exp \Big[
  iS_{\mathrm{eff}} [\pi_1,\pi_2;J_1,J_2] \Big]\, .
 \label{eq:CTPGF}
\end{align}
Here $\varphi$ denotes a dynamical degree of freedom in the UV theory, 
$\sigma$ a possible environment, $J$ an external source,
and $\rho_0$ an initial density operator. 
We have $\hat{U}_{J_1}$ and $\hat{U}^\dag_{J_2}$  
so that a number of fields is all doubled.  
The effective action for the doubled NG modes 
$S_{\mathrm{eff}} [\pi_1,\pi_2;J_1,J_2]$ can capture universal low-energy 
dynamics of systems with time-translational symmetry breaking 
in the similar way as the in-out formalism presented in the previous subsection.
However, we have to pay attention to a more complicated structure 
due to the doubling nature of the Schwinger-Keldysh formalism.

\medskip
Such a construction has been recently clarified in order to provide
the EFT for dissipative relativistic hydrodynamics~\cite{Crossley:2015evo,Glorioso:2017fpd}.
We briefly summarize basis of their formulation. 
The vital point is that we construct the effective action 
$S_{\mathrm{eff}} [\pi_1,\pi_2;J_1,J_2]$ respecting 
the following basic properties which the CTPGF satisfies 
(See Refs.~\cite{Crossley:2015evo,Glorioso:2017fpd} 
for derivation and discussion of these conditions in detail): 
\begin{enumerate}
 \item \textbf{Unitarity condition:}
       With a normalized initial density operator 
       $\mathrm{Tr}\,\hat{\rho}_0 = 1$,
       the CTPGF~\eqref{eq:CTPGF} satisfies 
       \begin{equation}
	Z[J_1 = J ,J_2 = J] = 1 ,
       \end{equation}
       from the unitarity of time evolution operator: 
       $\hat{U}_{J}^\dagger\hat{U}_{J}=1$ .
 \item \textbf{Conjugate condition:}
       The self-adjointness of the initial density operator $\hat{\rho}_{0}$ 
       ($\hat{\rho}_0^\dag = \hat{\rho}_0$) leads to 
       \begin{equation}
	Z^* [J_1, J_2] = Z[J_2, J_1] .
       \end{equation}
 \item \textbf{KMS condition (optional):}       
       When our initial density operator is thermal one 
       $\hat{\rho}_0 = e^{-\beta \hat{H}}/Z$, the following optional symmetry 
       emerges: 
       \begin{equation}
       \label{eq_KMS}
	Z [J_1,J_2 ] =  Z [J_1', J_2'] \with 
	 \begin{cases}
	  J_1'(t) = \epsilon_{\mathcal{T}_J} J_1 (-t + i \beta /2) , \\
	  J_2'(t) = \epsilon_{\mathcal{T}_J} J_2 (-t - i\beta/2) ,
	 \end{cases}
       \end{equation}
       where $\epsilon_{\mathcal{T}_J}$ denotes an eigenvalue of 
       the operator coupled to $J$ under 
       time-reversal (and simultaneous transformation of parity, if necessary).
\end{enumerate}
Note that while the first two conditions are quite general, 
the final one need a strong assumption on the initial density operator 
$\rho_0$, and thus, we treat it as optional.

\medskip
Then, the problem is how we constrain the structure of the EFT 
based on these properties. 
In order to safely satisfy the above conditions, we put some assumptions 
on the effective action $S_{\mathrm{eff}} [\pi_1,\pi_2;J_1,J_2]$. 
For simplicity, we will drop the external source from now on, but 
we can easily include them. 
The unitarity condition and the conjugate condition can then be implemented 
into $S_{\mathrm{eff}} [\pi_1,\pi_2]$ by imposing
\begin{align}
 S_{\mathrm{eff}} [\pi,\pi] &=0 ,\label{+-unitary:action}
 \\
 S_{\mathrm{eff}} [\pi_1,\pi_2]& = -S[\pi_2,\pi_1]^*,
\label{+-unitarity:2}
\end{align}
where $S[\pi_1,\pi_2]^*$ denote a complex conjugate of $S[\pi_1,\pi_2]$.
The second relation \eqref{+-unitarity:2} says that 
we generally have the imaginary part for the effective action. 
This is a basic feature of open systems since the dynamics after integrating 
out environment (UV) degrees of freedom contains fluctuation and dissipation 
in general.
Then, recalling that we have $e^{iS_{\mathrm{eff}}}$ for the weight of 
path integral, we may have divergent CTPGF 
if $\mathrm{Im} S_{\mathrm{eff}} < 0$.
To avoid such crisis, we also impose the positivity of the imaginary-part 
of the effective action%
\footnote{
See \cite{Glorioso:2016gsa} for the derivation of the positivity condition.
}:
\begin{align}
 \mathrm{Im} S_{\mathrm{eff}} [\pi_1,\pi_2] \geq 0.\label{unitary:open}
\end{align}
It is also convenient to introduce $RA$ basis 
defined by
\begin{align}
 \pi_R = \frac{\pi_1 + \pi_2}{2}\,,
 \quad
 \pi_A=\pi_1-\pi_2\,.
\label{def:RAbasis}
\end{align}
Then, the unitarity and conjugate condition 
for the effective action~(\ref{+-unitary:action}) and (\ref{+-unitarity:2}) 
can be rewritten as
\begin{align}
S[\pi_{R},\pi_{A}=0]&=0,\label{unitary:action}
\\
S[\pi_{R},\pi_{A}]&=-S [\pi_{R},-\pi_{A}]^*,
\label{unitarity:2}
\end{align}
which we will use in the next subsection.
We can also put a restriction on the EFT from the KMS condition.
However, since it may or may not emerge in the EFT, we will 
discuss this issue after constructing the general effective Lagrangian 
in the next subsection.

\subsection{Effective Lagrangian in open systems}
\label{subsec:EFT_const}

We then construct the EFT for the NG modes in open systems. As we elaborated in Sec.~\ref{micro:origin}, all the microscopic symmetries are doubled in the Schwinger-Keldysh formalism. Corresponding to the nonlinear transformation rules~\eqref{translations}-\eqref{Lorentz}, the doubled Poincar\'e symmetry transformations of the doubled NG modes are given by ($a=1,2$)
\begin{align}
\label{pi_a_translations}
 \pi_a(t,\mathbf{x})
 &\to \pi_a'(t,\mathbf{x})
 = \pi_a\left(t+\epsilon^0_a,\mathbf{x}+\boldsymbol{\epsilon}_a\right)+\epsilon^0_a\,,
\\
\pi_a(t,\mathbf{x})
 &\to \pi_a' (t,\mathbf{x}) 
 = \pi_a \left(\Lambda_a^0\,{}_\mu x^\mu,\Lambda_a^i\,{}_\mu x^\mu\right)+\Lambda_a^0\,{}_\mu x^\mu-t
\,,
\end{align}
where $\epsilon_a$ and $\Lambda_a^\mu\,{}_\nu\in SO(1,3)$ are transformation parameters for the doubled translations and Lorentz symmetries. 
In particular, the open system nature explicitly breaks the doubled symmetries into the diagonal ones: $\epsilon_1=\epsilon_2=\epsilon_R$ and $\Lambda_1=\Lambda_2=\Lambda_R$.
We call these two the $\epsilon_R$- and $\Lambda_R$-symmetries here and hereafter, while we called the $\epsilon_R$-symmetry the $H_A$ symmetry in the previous section. We also parameterize the broken symmetries as $\displaystyle\epsilon_1=-\epsilon_2=\frac{\epsilon_A}{2}$ and $\Lambda_1=(\Lambda_2)^{-1}=(\Lambda_A)^{1/2}$, and call them the $\epsilon_A$- and $\Lambda_A$-symmetries.
In the $RA$ basis~\eqref{def:RAbasis}, the $\epsilon_R$-transformations of NG modes are
\begin{align}
\label{R_translations_R}
 \pi_R(t,\mathbf{x})
 &\to \pi_R'(t,\mathbf{x})
 = \pi_R\left(t+\epsilon^0_R,\mathbf{x}+\boldsymbol{\epsilon}_R\right)+\epsilon^0_R\,,
\\
\label{R_translations_A}
  \pi_A(t,\mathbf{x})
 &\to \pi_A'(t,\mathbf{x})
 = \pi_A\left(t+\epsilon^0_R,\mathbf{x}+\boldsymbol{\epsilon}_R\right),
\end{align}
whereas the $\Lambda_R$-transformations are  given by
\begin{align}
\label{R_Lorentz_R}
\pi_R(t,\mathbf{x})
 &\to \pi_R' (t,\mathbf{x}) 
 = \pi_R \left(\Lambda_R^0\,{}_\mu x^\mu,\Lambda_R^i\,{}_\mu x^\mu\right)+\Lambda_R^0\,{}_\mu x^\mu-t\,,
 \\
\label{R_Lorentz_A}
 \pi_A(t,\mathbf{x})
 &\to \pi_A' (t,\mathbf{x}) 
 = \pi_A \left(\Lambda_R^0\,{}_\mu x^\mu,\Lambda_R^i\,{}_\mu x^\mu\right)\,.
\end{align}
An important point here is that the $\pi_A$ field {\it linearly} transforms under the $\epsilon_R$- and $\Lambda_R$-symmetries, just as an ordinary matter. 
In a similar way, we may see that the $\pi_{R/A}$ field linearly/nonlinearly transforms under the $\epsilon_A$- and $\Lambda_A$-symmetries (we will provide the transformation rule in the RA basis explicitly later when necessary).
In the following, we construct and analyze the effective Lagrangian of the doubled NG modes based on this symmetry structure and the Schwinger-Keldysh conditions summarized in Sec.~\ref{subsec:EFT_in-in}.

\subsubsection{General Lagrangian}

As we have just mentioned, the $\pi_A$ field can be thought of as a matter field from the $\epsilon_R$- and $\Lambda_R$-symmetry viewpoint. Therefore, the general ingredients for the open system effective action, which respects the $\epsilon_R$-and $\Lambda_R$-symmetry only, are given by
\begin{align}
\pi_{A},~t+\pi_{R} {\rm ~and~their~derivatives},
\label{ingredients}
\end{align}
which are invariant under Eqs.~\eqref{R_translations_R}-\eqref{R_Lorentz_A} upon an appropriate coordinate transformation, just as the in-out case. Based on these ingredients, we construct the general Lagrangian consistent with the positivity condition~\eqref{unitary:open}, the unitarity condition (\ref{unitary:action}) and the conjugate condition (\ref{unitarity:2}).

\paragraph{$\hbar$ expansion}
Let us first recall that the $\hbar$-dependence of $RA$ fields is given by
\begin{align}
\pi_R=\mathcal{O}(\hbar^0)\,,
\quad
\pi_A=\mathcal{O}(\hbar)\,.
\end{align}
The effective Lagrangian can then be expanded in $\hbar$ as
\begin{align}
\mathcal{L}_{\rm eff}=\sum_{n=1}^\infty \mathcal{L}_n
\quad
{\rm s.t.}
\quad
\mathcal{L}_n=\mathcal{O}(\pi_A^n)=\mathcal{O}(\hbar^n)\,,
\end{align}
where note that the $\hbar$ expansion corresponds to the expansion 
in the number of $\pi_A$ and the Lagrangian starts from the first order term 
in $\pi_A$ because of the unitarity condition~\eqref{unitary:action}. 
Also, the conjugate condition~(\ref{unitarity:2}) requires that the operators 
in $\mathcal{L}_n$ with an odd (even) $n$ have a real (pure imaginary) constant.
As we will see, the leading order Lagrangian $\mathcal{L}_1$ contains 
the dissipation term and the second order Lagrangian $\mathcal{L}_2$ contains 
the noise term. 
We call the truncation at $\mathcal{O}(\hbar^2)$
the semiclassical limit, which is
equivalent to working at the MSR action level. 
In the following, for technical simplicity, we work in this semiclassical 
limit and construct $\mathcal{L}_1$ and $\mathcal{L}_2$ only, even though 
it is straightforward to extend the construction to higher-order in $\hbar$.

\paragraph{An illustrative example for $\mathcal{L}_1$}

Let us start from the leading order $\mathcal{L}_1$. Before considering the general Lagrangian, it is illustrative to consider the following simple example:
\begin{align}
\mathcal{L}_1^{\rm simple}&=
\gamma_0(t+\pi_R)\pi_A
+\gamma_1(t+\pi_R)(P_\mu P^\mu+1)\pi_A
 -\alpha_1(t+\pi_R)P^\mu\partial_\mu\pi_A\,,
 \label{Eq:SimpleL}
\end{align}
where we introduced 
$P_\mu=\partial_\mu(t+\pi_R)=\delta_\mu^0+\partial_\mu\pi_R$. 
Just as the in-out case, the EFT parameters $\gamma_0$, $\gamma_1$ 
and $\alpha_1$ are functions of $t+\pi_R$, which have to be real 
because of the conjugate condition (\ref{unitarity:2}). 
Also, the Lagrangian contains terms linear in $\pi$, 
\begin{align}
\mathcal{L}_1^{\rm simple}&=\gamma_0(t)\pi_A+\alpha_1(t)\dot{\pi}_A+\mathcal{O}(\pi^2)\,,
\end{align}
so that we impose the background equation of motion for $\pi_A$
to require $\gamma_0=\alpha_1'$. The Lagrangian \eqref{Eq:SimpleL} is 
then reduced to
\begin{align}
\mathcal{L}_1^{\rm simple}
&=
\gamma_1(t+\pi_R)\left(-2\dot{\pi}_R+(\partial_\mu\pi_R)^2\right)\pi_A
-\alpha_1(t+\pi_R)\partial_\mu\pi_R\partial^\mu\pi_A
-\alpha_1'(t+\pi_R)\dot{\pi}_R\pi_A\,,
\end{align}
where we dropped total derivatives. 
For intuitive understanding of these operators, it is useful to consider the slow-roll regime, where the EFT parameters $\alpha_1$ and $\gamma_1$ are constant:
\begin{align}
\mathcal{L}_1^{\rm simple}
&=
\gamma_1\left(-2\dot{\pi}_R+(\partial_\mu\pi_R)^2\right)\pi_A
-\alpha_1\partial_\mu\pi_R\partial^\mu\pi_A\,.
\end{align}
As is understood, e.g., from the comparison to Eq.~\eqref{eq:MSR2}, the $\alpha_1$ term is an ordinary kinetic term of the NG boson $\pi$. On the other hand, the $\gamma_1$ term contains a dissipation term $\dot{\pi}_R\pi_A$. Interestingly, in the time-translational symmetry broken phase of open systems, the dissipation term may appear without spoiling the boost symmetry, which is nonlinearly realized by accompanying the cubic interaction term $(\partial_\mu\pi_R)^2\pi_A$.

\paragraph{Construction of general $\mathcal{L}_1$}

We then consider the general $\mathcal{L}_1$. 
Similarly to the in-out case~\eqref{S_in-out_P}, 
let us restrict ourselves to operators containing $\pi_{R/A}$
with at most one derivative. Under this assumption, the general Lagrangian is given by
\begin{align}
\mathcal{L}_1&=
\mathcal{L}_1^{\rm simple}+\sum_{n=2}^\infty\gamma_n(P^\mu P_\mu+1)^n\pi_A
-\sum_{n=2}^\infty\alpha_n(P^\mu P_\mu+1)^{n-1}P^\mu\partial_\mu\pi_A\,,
\end{align}
where $\gamma_n$'s and $\alpha_n$'s are real functions of $t+\pi_R$. Notice that the terms linear in $\pi$ appear only in $\mathcal{L}_1^{\rm simple}$, so that the background equation of motion is the same as before. More explicitly, we have
\begin{align}
\nonumber
\mathcal{L}_1&=
-\alpha_1(t+\pi_R)\partial_\mu\pi_R\partial^\mu\pi_A
-\alpha_1'(t+\pi_R)\dot{\pi}_R\pi_A
\\
\nonumber
&\quad
-\sum_{n=2}^\infty\alpha_n(t+\pi_R)\left[-2\dot{\pi}_R+(\partial_\mu\pi_R)^2\right]^{n-1}(-\dot{\pi}_A+\partial_\mu\pi_R\partial^\mu\pi_A)
\\
&\quad
+\sum_{n=1}^\infty\gamma_n(t+\pi_R)\left[-2\dot{\pi}_R+(\partial_\mu\pi_R)^2\right]^n\pi_A\,.
\label{EFTLag:pi_A1}
\end{align}
In particular, only the three operators $\alpha_1$, $\alpha_2$, and $\gamma_1$ provide quadratic terms in $\pi$, which are relevant to the dispersion relation
of the NG mode.

\paragraph{Construction of $\mathcal{L}_2$}

We then move on to the second order term $\mathcal{L}_2$. 
Let us again focus on the operators containing $\pi_{R/A}$ with 
at most one derivative. 
Under this assumption, there are four operators relevant to the dispersion relation:
\begin{align}
\label{L2_simple_P}
\mathcal{L}_2\ni i\Big[
\beta_1\pi_A^2+\beta_2(\partial_\mu\pi_A)^2
+\beta_3(P^\mu\partial_\mu\pi_A)\,\pi_A
+\beta_4(P^\mu\partial_\mu\pi_A)^2
\Big]\,,
\end{align}
where $\beta_i$'s are real function of $t+\pi_R$.
More explicitly, we write 
\begin{align}
\nonumber
\mathcal{L}_2&\ni i\Big[
\beta_1(t+\pi_R)\pi_A^2+\beta_2(t+\pi_R)(\partial_\mu\pi_A)^2
+\beta_3(t+\pi_R)(-\dot{\pi}_A+\partial_\mu\pi_R\partial^\mu\pi_A)\pi_A
\\
\nonumber
&\qquad
+\beta_4(t+\pi_R)\big(\dot{\pi}_A^2-2\dot{\pi}_A\partial_\mu\pi_R\partial^\mu\pi_A
+(\partial_\mu\pi_R\partial^\mu\pi_A)^2\big)
\Big]
\\
\label{L2_simple}
&= i\Big[
\big(\beta_1(t) + \tfrac{1}{2}\beta_3'(t)\big)\pi_A^2+\beta_2(t)(\partial_\mu\pi_A)^2
+\beta_4(t)\dot{\pi}_A^2
+\mathcal{O}(\pi^3)\Big]
\,,
\end{align}
where we dropped total derivatives at the equality. 
Also notice that the $\beta_1$ and $\beta_3$ operators are degenerate at the quadratic level in $\pi$, even though they provide independent higher-order interaction terms.

\medskip
Among the three terms in the last line of Eq.~\eqref{L2_simple}, 
the first term is the ordinary noise term with a time-dependent coefficient. 
On the other hand, the other two terms are higher derivative corrections, 
which make the noise scale-dependent. 
In particular, the $\beta_4$ term breaks the Lorentz symmetry, 
so that it has to be accompanied by the cubic and quartic interaction terms 
to nonlinearly realize the spontaneously broken boost symmetry.
In a similar way, the general operators in $\mathcal{L}_2$ can be obtained 
by multiplying an arbitrary power of 
$(P_\mu P^\mu+1)=-2\dot{\pi}_R+(\partial_\mu\pi_R)^2$ to the four operators 
displayed above. 
These new operators generate cubic and higher-order interaction terms, 
hence they are not relevant to the dispersion relation and also subleading 
in the weakly coupled regime.

\medskip
As we mentioned, the conjugate condition~(\ref{unitarity:2}) requires that $\mathcal{L}_n$ with an even $n$ contributes to the imaginary part of the effective action, whose sign is fixed by the positivity condition~\eqref{unitary:open}. In the weakly coupled regime, the quadratic terms dominate over the cubic and higher-order interactions, so that the leading contribution to the imaginary part is the three quadratic terms in Eq.~\eqref{L2_simple}. Notice here that $\mathcal{L}_n$ ($n\geq3$) does not provide any quadratic term because it contains more than two $\pi_A$'s by definition. Also, as long as the derivative expansion works, the first term dominates over the other terms. Under these assumptions, the positivity condition~\eqref{unitary:open} can be stated as $\beta_1-\tfrac{1}{2}\beta_3'>0$.

\subsubsection{Low-energy spectrum}

We then discuss the dispersion relation in the low-energy regime.
To determine the dispersion relation, 
let us focus on the slow-roll regime, where the EFT parameters are treated as 
constant and the energy is well-defined. 
In this regime, the quadratic part of the effective Lagrangian takes the form,
\begin{align}
\nonumber
\mathcal{L}_{\rm eff}&\ni
(\alpha_1-2\alpha_2)\dot{\pi}_R\dot{\pi}_A-\alpha_1\partial_i\pi_R\partial_i\pi_A
-2\gamma_1\dot{\pi}_R\pi_A
\\
\label{quadratic_EFT}
&\quad
+i\Big[
\beta_1\pi_A^2-(\beta_2-\beta_4)\dot{\pi}_A^2+\beta_2(\partial_i\pi_A)^2
\Big]\,,
\end{align}
where all the EFT parameters are real constants and $\beta_1>0$ is required by the positivity condition~\eqref{unitary:open}.  In the low-energy limit, the following operators dominate over the others:
\begin{align}
\nonumber
\mathcal{L}_{\rm eff}&\ni
-\alpha_1\partial_i\pi_R\partial_i\pi_A
-2\gamma_1\dot{\pi}_R\pi_A
+i\beta_1\pi_A^2\,,
\end{align}
where we used
\begin{align}
\label{low-energy_assumptions}
\omega\ll\gamma_1/\alpha_i\,,
\quad
\omega^2,k^2\ll|\beta_1/\beta_{2,4}|\,,
\end{align}
to drop higher-derivative terms. We will provide a physical interpretation of these two conditions later, but they are satisfied in any case as long as we consider a sufficiently low-energy scale. In this low-energy limit, the on-shell condition can then be stated as
\begin{align}
{\rm det}\left(
\begin{array}{cc}0 &-\frac{1}{2} \alpha_1k^2- i\gamma_1\omega \\ -\frac{1}{2}\alpha_1k^2+i\gamma_1\omega & i\beta_1\end{array}
\right)=0
\quad
\Leftrightarrow
\quad
\omega^2=-\frac{\alpha_1^2}{4\gamma_1^2}k^4\,.
\end{align}
Here note that the noise term $\beta_1$ does not affect 
the dispersion relation. 
Interestingly, we find that the doubled NG modes form a canonical pair and 
describe a single {\it diffusive} mode with a quadratic dispersion. 
This is in a sharp contrast to the NG modes in closed systems. 
Such a dispersion relation of NG modes in open systems was found earlier 
in the case of internal symmetry breaking~\cite{Minami:2018oxl}. 
The origin of diffusive modes and quadratic dispersion in our setup is essentially the same as the internal symmetry breaking case discussed there.

\subsubsection{Energy scales}

Now let us get back to the two conditions~\eqref{low-energy_assumptions} 
which we used to take the low-energy limit. 
First, the second condition is easy to understand: 
it simply requires that the derivative expansion works among the operators generating the noise effects. On the other hand, the first condition needs some more consideration: As we explained, the $\gamma_1$ operator is the dissipation term, whereas the $\alpha_1$ operator is an ordinary kinetic term. Therefore, the first condition means that the dissipation term dominates over the temporal kinetic term. In other words, this condition characterizes the energy scale $E_{\rm diss}$ of the dissipation effects as
\begin{align}
E_{\rm diss}\sim\gamma_1/\alpha_1\,.
\end{align}
The low-energy limit discussed above may then be stated as $\omega\ll E_{\rm diss}$. Let us here recall that the time-translational symmetry breaking scale $E_{\rm SSB}$ is characterized as
\begin{align}
E_{\rm SSB}^4\sim \alpha_1\,.
\end{align}
Since the dissipative effects and the symmetry breaking have different origins, the two scales $E_{\rm diss}$ and $E_{\rm SSB}$ are generally independent. In particular, if there exists a hierarchy $E_{\rm diss}\ll E_{\rm SSB}$, there is a scale $\omega$ satisfying $E_{\rm diss}\ll \omega\ll E_{\rm SSB}$, where the dissipation effects are subleading contributions to the NG mode dynamics. Therefore, if we go beyond the low-energy limit, it becomes important to clarify which operators are associated with the dissipation effects and more generally specific to open systems~\footnote{For the opposite hierarchy, $E_{\rm SSB}\ll \omega\ll E_{\rm diss}$, the EFT description is no longer applicable, and there simply exist fluctuation and dissipation in the original UV theory.}.

\medskip
Based on this motivation, let us classify our EFT operators. First, $\mathcal{L}_1$ and $\mathcal{L}_2$ are at different orders in the $\hbar$ expansion. As is understood from the fact that $\mathcal{L}_2$ is pure imaginary, it describes the statistical noise, so that it is specific to open systems.
On the other hand, $\mathcal{L}_1$ contains both of operators specific to open systems and those which may exist also in closed systems. As we discussed in Sec.~\ref{micro:origin}, the latter operators enjoy the $\epsilon_A$- and $\Lambda_A$-symmetry. In particular, the $\epsilon_A$ time-translational symmetry transformations of the doubled NG modes are given by setting $\epsilon^0_1=-\epsilon^0_2=\epsilon_A/2$ in Eq.~\eqref{pi_a_translations} as
\begin{align}
\label{epsilon_A_R}
 \pi_R(t,\mathbf{x})&\to 
 \pi'_R(t,\mathbf{x}) 
 = \pi_R(t,\mathbf{x}) 
 +\mathcal{O}(\hbar^2)\,,
 \\
\label{epsilon_A_A}
 \pi_A(t,\mathbf{x})&\to 
 \pi'_A(t,\mathbf{x})
 =\pi_A(t,\mathbf{x})+\dot{\pi}_R(t,\mathbf{x})\epsilon_A+\epsilon_A+ O(\hbar^3)\,,
\end{align}
where we assigned an $\hbar$-dependence of the transformation parameter $\epsilon_A$ as $\epsilon_A=\mathcal{O}(\hbar)$. Notice that the truncation at this order is consistent with the semiclassical limit of the Schwinger-Keldysh action,
i.e., the MSR effective action.
We then consider the $\epsilon_A$ transformation of $\mathcal{L}_1$. For simplicity, let us focus on the slow-roll regime:
\begin{align}
\mathcal{L}_1&=
\sum_{n=1}^\infty\gamma_n(P^\mu P_\mu+1)^n\pi_A
-\sum_{n=1}^\infty\alpha_n(P^\mu P_\mu+1)^{n-1}P^\mu\partial_\mu\pi_A
\,,
\end{align}
where $\alpha_n$'s and $\gamma_n$'s are constants. 
By noting the following transformation property,
\begin{align}
\delta_{\epsilon_A} (P^\mu\partial_\mu\pi_A)=\epsilon_A\,P^\mu\partial_\mu(\dot{\pi}_R+1)=\frac{1}{2}\epsilon_A\,\partial_t(P^\mu P_\mu+1)\,,
\end{align}
the $\epsilon_A$-transformation of $\mathcal{L}_1$ can be calculated as
\begin{align}
\delta_{\epsilon_A} \mathcal{L}_1&=
\epsilon_A\left[
\sum_{n=1}^\infty\gamma_n(P^\mu P_\mu+1)^n(\dot{\pi}_R+1)
-\sum_{n=1}^\infty\alpha_n\frac{1}{2n}\partial_t(P^\mu P_\mu+1)^{n}
\right]
\,.
\end{align}
We find that the $\alpha_n$ operators are invariant under the $\epsilon_A$ time-translation because the second term is a total derivative\footnote{
\label{footnote:non-slow_roll}
If we relax the slow-roll assumption, the $\epsilon_A$-invariant operators are given by the choice $\alpha_n'= - 2n\gamma_n$.}. We may also show that they enjoy the $\Lambda_A$-symmetry as well. Therefore, the $\alpha_n$ operators may exist even in closed systems, while the $\gamma_n$ operators are specific to open systems. The corresponding EFT coefficients are then estimated as
\begin{align}
\alpha_n\sim E_{\rm SSB}^4\,,
\quad
\gamma_n\sim E_{\rm diss}\,E_{\rm SSB}^4\,.
\end{align}

\medskip
Finally, we derive the dispersion relation valid beyond the low-energy limit. Suppose that there is a hierarchy $E_{\rm diss}\ll E_{\rm SSB}$ and there exists an intermediate scale $E_{\rm diss}\lesssim \omega\ll E_{\rm SSB}$, where the low-energy limit result is no more applicable. In this intermediate scale, the following operators may be the leading operators in the quadratic Lagrangian~\eqref{quadratic_EFT}:
\begin{align}
\nonumber
\mathcal{L}_{\rm eff}&\ni
(\alpha_1-2\alpha_2)\dot{\pi}_R\dot{\pi}_A-\alpha_1\partial_i\pi_R\partial_i\pi_A
-2\gamma_1\dot{\pi}_R\pi_A+i\beta_1\pi_A^2
\\
&=
(\alpha_1-2\alpha_2)\left(
\dot{\pi}_R\dot{\pi}_A-c_s^2\partial_i\pi_R\partial_i\pi_A
-\gamma\dot{\pi}_R\pi_A+i\frac{A}{2}\pi_A^2
\right)
\,,
\label{eq:lag_eff_intscale}
\end{align}
where we dropped higher derivative corrections to the noise terms. We also introduced
\begin{align}
c_s^2=\frac{\alpha_1}{\alpha_1-2\alpha_2}\,,
\quad
\gamma=\frac{2\gamma_1}{\alpha_1-2\alpha_2}\,,
\quad
A=\frac{2\beta_1}{\alpha_1-2\alpha_2}\,,
\end{align}
where $\gamma\sim E_{\rm diss}$ and $A$ are the damping coefficient and the noise amplitude, respectively. 
Also, $c_s$ denotes the propagation speed of the NG mode.
The dispersion relation is then
\begin{align}
{\rm det}\left(
\begin{array}{cc}0 &\omega^2-c_s^2k^2- i\gamma\omega \\ \omega^2-c_s^2k^2+ i\gamma\omega & iA\end{array}
\right)=0
\quad
\Leftrightarrow
\quad
\omega^2=c_{s}^{2}k^2-\frac{\gamma^2}{2}\pm\sqrt{\frac{\gamma^4}{4}-\gamma^2c_{s}^{2}k^2}\,.
\end{align}
In the low-energy limit $c_{s}k\ll\gamma\sim E_{\rm diss}$, we find one gapless and one gapped diffusive modes:
\begin{align}
\omega^2\simeq- \frac{c_{s}^{4}k^4}{\gamma^2}\,,
\quad
\omega^2\simeq-  \gamma^2+2c_{s}^{2}k^2\,,
\end{align}
where note that the gapped mode was not captured in the previous argument because the temporal kinetic term $\dot{\pi}_R\dot{\pi}_A$ was neglected by taking the low-energy limit. 
On the other hand, at the short-length scale satisfying
$c_{s}k\gg \gamma\sim E_{\rm diss}$,  
we find two propagating modes with small dissipations:
\begin{align}
\omega^2\simeq c_{s}^{2}k^2\pm i \gamma c_{s}k\,.
\end{align}

\subsubsection{Restriction to EFT from the dynamical KMS symmetry}
\label{subsec:KMS}

As we discussed in Sec.~\ref{subsec:KMS_B}, an additional discrete symmetry called the KMS symmetry emerges when our system initially stays in a thermal equilibrium state. In the rest of this subsection we discuss its implication for the low-energy coefficients of the effective action.

\paragraph{KMS transformation}

\medskip
In Sec.~\ref{subsec:KMS_B} we introduced the dynamical KMS transformation as
a combination of the $\epsilon_A$ time-translation~\eqref{epsilon_A_R}-\eqref{epsilon_A_A} with a pure-imaginary transformation parameter $\epsilon_A=-i\beta$ ($\beta$ is the inverse temperature) and the time-reversal transformation. To identify the KMS transformation of the NG fields, it is convenient to introduce a condensation field $\phi$ with a time-dependent background:
\begin{align}
\langle\phi_a(t,{\bf x})\rangle=\bar{\phi}(t)\,,
\end{align}
where $a=1,2$ is the label of the doubled fields on the Keldysh contour. The double NG fields $\pi_a$ ($a=1,2$) may then be embedded as
\begin{align}
\phi_a(t,{\bf x})=\bar\phi(t+\pi_a(t,\mathbf x))\,.
\end{align}
If the condensation field $\phi$ has an even time-reversal parity,
\begin{align}
\phi_a(t,{\bf x})\to\phi_a'(t,{\bf x})=\phi_a(-t,{\bf x})
\quad
(a=1,2)\,,
\label{ota}
\end{align}
the time-reversal transformation of the NG fields is given by
\begin{align}
\pi_a(t,\mathbf x)\to \pi_a'(t,\mathbf x) = -2t+\pi_a(-t,\mathbf x)\quad
(a=1,2)\,.
\end{align}
In the RA basis we may rephrase it as
\begin{align}
    \pi'_R(t,\mathbf x)=-2t+\pi_R(-t,\mathbf x),~~
        \pi'_A(t,\mathbf x)=\pi_A(-t,\mathbf x).
        \label{time:rev}
\end{align}
Note that the time-reversal symmetry is nonlinearly realized by the NG fields, essentially because the time-dependent background generically breaks the time-reversal symmetry.

\medskip
One can accordingly introduce a dynamical KMS transformation, using the above time-reversal transformation with the $\epsilon_A$ time-translation. 
We again take the semiclassical limit and work at the MSR action level. 
Using Eqs.~\eqref{epsilon_A_R}-\eqref{epsilon_A_A} with the parameter $\epsilon_A=-i\beta$ and Eq.~\eqref{time:rev}, we obtain the KMS transformation of the form,
\begin{align}
\label{KMS}
 \pi_R'(t,\mathbf x) =-2t+\pi_R(-t,\mathbf x)
 \,,
 \quad
 \pi_A'(t,\mathbf x) = 
 \pi_A(-t,\mathbf x) 
 + i\beta \big[\partial_{-t} \pi_R(-t,\mathbf x)+1\big]\,,
\end{align}
where notice that both of $\pi_R$ and $\pi_A$ are nonlinearly transformed
\footnote{
In the context of dissipative fluids~\cite{Glorioso:2017fpd} a linear KMS transformation rule,
\begin{align}
\label{dfKMS}
\pi_R'(t,\mathbf x) =-\pi_R(-t,\mathbf x)
 \,,
 \quad
 \pi_A'(t,\mathbf x) = 
 -\pi_A(-t,\mathbf x) 
 - i\beta \partial_{-t} \pi_R(-t,\mathbf x)\,,
\end{align}
is often employed rather than the nonlinear one~\eqref{KMS}. In this context, the condensation field $\phi$ has an odd parity and has a slow-roll type background $\langle\phi_a(x)\rangle=vt$ with a constant $v$. From the embedding $\phi_a(t,\mathbf x)=v(t+\pi_a(t,\mathbf x))$ and the time-reversal transformation $\phi_a(t,{\bf x})\to\phi_a'(t,{\bf x})=-\phi_a(-t,{\bf x})$, the dynamical KMS transformation rule of the NG fields follows as
\begin{align}
\label{dfKMS_pre}
 \pi_R'(t,\mathbf x) =-\pi_R(-t,\mathbf x)
 \,,
 \quad
 \pi_A'(t,\mathbf x) = 
 -\pi_A(-t,\mathbf x) 
 - i\beta \big[\partial_{-t} \pi_R(-t,\mathbf x)+1\big]\,,
\end{align}
where note that $\pi_R$ transforms linearly, but $\pi_A$ nonlinearly at this stage. Furthermore, in dissipative fluids, the shift symmetry of $\pi_A$ is imposed to realize the energy conservation of the full system. As a result, the nonlinear transformation~\eqref{dfKMS_pre} may be reduced to the linear one~\eqref{dfKMS} accompanied by an appropriate constant shift of $\pi_A$. In this way, the existence of extra symmetries is crucial to have a linearly realized KMS symmetry. Since our paper is considering more generic setups for time-translational symmetry breaking, we employed the nonlinearly realized one~\eqref{KMS} in contrast to the dissipative fluid case.}.

\paragraph{Constraints on the quadratic Lagrangian}

We then discuss an optional constraint resulting from invariance under the dynamical KMS transformation. For illustration, we focus on the slow-roll regime, where the time-dependence of the EFT coefficients, $\alpha_n, \beta_n, \gamma_n$, is negligible, and demonstrate how the KMS invariance constrain the EFT parameters.

\medskip
Let us start with the KMS transformation of 
the second-order Lagrangian~\eqref{eq:lag_eff_intscale}. 
First, it is easy to show that the $\alpha_n$ operators are invariant under the KMS transformations upon a coordinate transformation $t\to-t$. 
Under the KMS transformation and a coordinate change $t\to-t$, the dissipation term is transformed as 
\begin{align}
 -\gamma\dot{\pi}_R\pi_A 
 &\to \gamma\dot{\pi}_R\pi_A 
 +2\gamma\pi_A+2i\gamma\beta(\dot{\pi}_R+1)
 + i\gamma\beta\dot{\pi}_R(\dot{\pi}_R+1)\,,
\end{align}
whereas the fluctuation term is transformed as
\begin{align}
 i\frac{A}{2}\pi_A^2 
 \to i\frac{A}{2}\pi_A^2 
-A\beta(\dot{\pi}_R+1)\pi_A
 - i\frac{A\beta^2}{2}(\dot{\pi}_R+1)^2\,.
\end{align}
Therefore, the second-order Lagrangian~\eqref{eq:lag_eff_intscale} becomes 
invariant under the KMS transformations~\eqref{KMS} up to total derivatives,
if the damping coefficient $\gamma$ and the noise amplitude $A$ satisfy the relation:
\begin{align}
A=\frac{2\gamma}{\beta}\,,
\end{align}
which is nothing but the fluctuation-dissipation relation (FDR). 
Since $A>0$ is required by the positivity condition~\eqref{unitary:open}, it turns out that the damping coefficient is positive $\gamma>0$ if the FDR is satisfied.

\paragraph{Beyond the quadratic level}

We then incorporate the interaction terms. A nontrivial point here is that the dissipation term $\dot{\pi}_R\pi_A$ has to be accompanied by the cubic interaction $\pi_A(\partial_\mu\pi_R)^2$ as long as we respect the (nonlinearly realized) boost symmetry. It is easy to see that this cubic term $\pi_A(\partial_\mu\pi_R)^2$ is not invariant under the KMS transformation, even if the FDR is satisfied. As long as we know, there is no set of EFT parameters which respects both of the boost and the dynamical KMS symmetries. This is not so surprising because the finite temperature effects break the boost symmetry. Indeed, if we give up the boost symmetry (while respecting the nonlinearly realized time-translation), we may use $\delta^0_\mu$ in the construction of effective Lagrangian. For example, we may introduce the dissipation term,
\begin{align}
\label{Lorentz_breaking_dissipation}
\widetilde{\gamma}_1\left[\delta^0_\mu\partial^\mu(t+\pi_R)\pi_A+\pi_A\right]=-\widetilde{\gamma}_1\dot{\pi}_R\pi_A\,,
\end{align}
without cubic interactions (we denoted the EFT coefficient by $\widetilde{\gamma}_1$). This operator gives a dynamical KMS invariant Lagrangian with the FDR.

\subsection{Model analysis}
\label{sec:ModelAnalysis}

\paragraph{The model}

At the end of this section, we consider a simple UV model composed of a single-component scalar $\phi$ to illustrate the relation between the low-energy coefficients of the EFT and information on the UV theory.
Taking into account a possible environment coupled to the scalar $\phi$, 
we start with the following Schwinger-Keldysh action in the semiclassical limit:
\begin{align}
\label{scalar_model}
S[\phi_R,\phi_A] 
 =\int d^4x\,\left[
\phi_A\Big(\Box\phi_R-V'(\phi_R)
 - \gamma\partial_t\phi_R\Big)
 +\frac{iA}{2}\phi_A^2
 \right]\,,
\end{align}
where the first two terms correspond to the closed system action with a canonical kinetic term and a potential $V(\phi)$. The last two terms denote the noise and dissipation terms. This model accommodates essentially the same symmetry structure 
as the Brownian particle system in the previous section: For arbitrary values of $\gamma$ and $A$, the action enjoys the $\epsilon_R$ time-translational symmetry,
\begin{align}
\nonumber
 &\phi_R (t,\mathbf{x}) 
 \xrightarrow { \epsilon_R }  \phi'_R(t,\mathbf{x}) = \phi_R (t+\epsilon_R,\mathbf{x})
 \,,
 \\
&\phi_A (t,\mathbf{x})
 \xrightarrow { \epsilon_R } \phi'_A (t,\mathbf{x}) = \phi_A (t+\epsilon_R,\mathbf{x})\,.
\end{align}
If there are no noise and dissipation, i.e., $\gamma=A=0$, there exists a symmetry enhancement, and the action also enjoys symmetry under 
the $\epsilon_A$ time-translation given by 
\begin{align}
 \nonumber
 &\phi_R(t,\mathbf{x})
 \xrightarrow { \epsilon_A } \phi'_R(t,\mathbf{x}) = \phi_R(t,\mathbf{x}) 
 \,,
 \\
 &\phi_A (t,\mathbf{x})
 \xrightarrow { \epsilon_A } \phi'_A (t,\mathbf{x}) = \phi_A (t,\mathbf{x})+\dot{\phi}_R (t,\mathbf{x})\epsilon_A  \,.
\end{align}
Note that the Lorentz symmetry is explicitly broken by the dissipation term.

\paragraph{Symmetry breaking}
We then discuss time-translational symmetry breaking in this model
and derive the effective Lagrangian for the NG fields.
Let us suppose that the scalar field has a time-dependent background,
\begin{align}
\langle\phi_R (t,\mathbf{x})\rangle
 =\bar{\phi}(t)\,, \quad \langle\phi_A (t,\mathbf{x})\rangle=0\,,
\end{align}
where $\bar{\phi}(t)$ is a spatially homogeneous solution 
of the equation of motion,
\begin{align}
\label{EMO:phibar}
\ddot{\bar{\phi}}+\gamma\dot{\bar{\phi}}+V'(\bar{\phi})=0\,.
\end{align}
The background then breaks both of the $\epsilon_R$ and $\epsilon_A$ time-translational symmetries. More explicitly, the background of $\phi_R$ transforms under the $\epsilon_R$-transformation as
\begin{align}
\label{transf_backgrounds}
 \langle\phi_R (t,\mathbf{x})\rangle 
 = \bar{\phi}(t) &\xrightarrow { \epsilon_R } \langle \phi'_R (t,\mathbf{x})\rangle =
 \langle \phi_R(t+\epsilon_R,\mathbf{x})\rangle=\bar{\phi}(t+\epsilon_R) 
 \,,
\end{align}
and the background of $\phi_A$ transforms under the $\epsilon_A$-transformation as
\begin{align}
 \langle\phi_A (t,\mathbf{x})\rangle 
 = 0  &\xrightarrow { \epsilon_A } \langle \phi'_A (t,\mathbf{x})\rangle =
 \langle \phi_A(t,\mathbf{x})+\dot{\phi}_R (t,\mathbf{x})\epsilon_A\rangle=\dot{\bar{\phi}}(t)\epsilon_A\,.
\end{align}
We therefore have two NG fields if $\gamma=A=0$, 
while one of the two becomes a pseudo NG field 
in the presence of fluctuation and dissipation.
By promoting the transformation parameters $\epsilon_R$ and $\epsilon_A$ 
in Eq.~\eqref{transf_backgrounds} to local fields $\pi_R (x)$ and $\pi_A (x)$, 
NG fields can be embedded into the original fields $\phi_R$ and $\phi_A$ as
\begin{align}
\label{phi_piRA}
 \phi_R(t,\mathbf{x}) 
 = \bar{\phi} \big( t+\pi_R(t,\mathbf{x}) \big)\,,
 \quad
 \phi_A(t,\mathbf{x}) 
 = \dot{\bar{\phi}} \big(t+\pi_R(t,\mathbf{x}) \big) \pi_A (t,\mathbf{x}) \,.
\end{align}
As we already discussed in the previous subsection, 
the transformation rule of the two NG fields is 
given by Eqs.~\eqref{R_translations_R}-\eqref{R_translations_A} 
and Eqs.~\eqref{epsilon_A_R}-\eqref{epsilon_A_A}.

\paragraph{Action of the NG fields}
The action for the NG fields can be obtained by substituting 
the relations~\eqref{phi_piRA} into the original action~\eqref{scalar_model}:
\begin{align}
S[\phi_R,\phi_A] 
=\int d^4x\left[
\dot{\bar{\phi}}\,\pi_A\Big(\dot{\bar{\phi}}\,\Box\pi_R+\ddot{\bar{\phi}}\,(-2\dot{\pi}_R+\partial_\mu\pi_R\partial^\mu\pi_R)
 - \gamma\dot{\bar{\phi}}\,\dot{\pi}_R\Big)
 +\frac{iA}{2}\dot{\bar{\phi}}^2\pi_A^2
 \right]\,,
\end{align}
where the arguments of $\bar{\phi}$ and derivatives are $t+\pi_R$. We also used the equation of motion~\eqref{EMO:phibar}. By performing a partial integral, we may rewrite it as
\begin{align}
\label{model_pi}
S[\phi_R,\phi_A] 
=\int d^4x\left[
-\dot{\bar{\phi}}^2
\partial_\mu\pi_R\partial^\mu\pi_A
-\ddot{\bar{\phi}}\dot{\bar{\phi}}\,
(\partial_\mu\pi_R\partial^\mu\pi_R)\pi_A
 - \gamma\dot{\bar{\phi}}^2\dot{\pi}_R\pi_A
 +\frac{iA}{2}\dot{\bar{\phi}}^2\pi_A^2
 \right]\,.
\end{align}
As we mentioned, our original setup~\eqref{scalar_model} explicitly breaks the Lorentz symmetry. We therefore need to add a Lorentz symmetry breaking operator~\eqref{Lorentz_breaking_dissipation}
into the effective Lagrangian~\eqref{EFTLag:pi_A1}-\eqref{L2_simple_P} in order to embed our model into the EFT framework:
\begin{align}
\label{EFT_Lbreaking}
\mathcal{L}_{\rm eff}=
-\alpha_1
\partial_\mu\pi_R\partial^\mu\pi_A
-\alpha_1'\dot{\pi}_R\pi_A
+\gamma_1
(-2\dot{\pi}_R+\partial_\mu\pi_R\partial^\mu\pi_R)\pi_A
+i\beta_1\pi_A^2
- \widetilde{\gamma}_1\dot{\pi}_R\pi_A\,,
\end{align}
where the first three terms are the leading order terms in the $\alpha_n$, $\gamma_n$, and $\beta_n$ sectors and the last term is the Lorentz symmetry breaking operator~\eqref{Lorentz_breaking_dissipation} introduced in Sec.~\ref{subsec:KMS}.
By comparing Eq.~\eqref{model_pi} and Eq.~\eqref{EFT_Lbreaking}, the low-energy coefficients read
\begin{align}
\nonumber
\alpha_1(t+\pi_R)=\dot{\bar{\phi}}^2(t+\pi_R),& \quad \widetilde{\gamma}_1(t+\pi_R)=\gamma\dot{\bar{\phi}}^2(t+\pi_R),
\\
\gamma_1(t+\pi_R)=-\dot{\bar{\phi}}\ddot{\bar{\phi}}(t+\pi_R),& \quad \beta_1(t+\pi_R)=\frac{iA}{2}\dot{\bar{\phi}}^2(t+\pi_R).
\end{align}
In particular, we observe that the coefficient $\alpha_1$ of the kinetic term is directly related to the order parameter $\dot{\bar{\phi}}$. 
We also find that this model satisfies $\alpha_1'=-2\gamma_1$, so that the first three terms in Eq.~\eqref{EFT_Lbreaking} are invariant under the $\epsilon_A$- and $\Lambda_A$-symmetries. (see footnote~\ref{footnote:non-slow_roll}).  
The derivation of the dispersion relation and the fluctuation-dissipation relation with dynamical KMS symmetry can be performed in the same way as the previous subsection.

\section{Summary and Outlook}
\label{section:Summary}

In this paper, we formulated a general way to construct the effective field 
theory associated with time-translational symmetry breaking for 
nonequilibrium open systems. 
After introducing basic concepts such as a \textit{weak} criterion of 
time-translational SSB by using the simplest example of the Brownian motion, 
we laid out a solid basis to construct the EFT for general situations
based on the doubled time-translational symmetry structure 
in the Schwinger-Keldysh formalism. 
The resulting EFT enables us to obtain the dispersion relations for the corresponding NG mode 
and gapped mode for open systems.
After constructing the most general effective Lagrangian, 
we also discussed a nontrivial restriction to low-energy (Wilson) coefficients 
coming from the dynamical KMS symmetry, 
which is regarded as a remnant of thermal properties of systems. 

\medskip
There are diverse nonequilibrium systems---in cosmology, 
condensed-matter physics and chemical and possibly biological or economic systems---where 
our formulation is applicable.
One promising application is to construct 
the open system EFT for the inflation in the early universe. 
Even though we focus on the flat space dynamics in this paper, 
it will be straightforward to extend our argument to curved spacetimes.
It will be useful to probe the hidden sector particles during the inflation epoch as a complementary approach to the so-called cosmological collider physics program~\cite{Chen:2009zp,Baumann:2011nk,Noumi:2012vr,Arkani-Hamed:2015bza}. It will also provide a model independent framework, e.g., for the stochastic inflation~\cite{stochastic} and the warm inflation~\cite{Berera:1995ie}.

\medskip
Another interesting direction is the application to condensed-matter physics
such as the cold-atomic systems. 
In fact, it has been recently pointed out that there exist 
a nonequilibrium phase transition and corresponding novel symmetry broken 
phase in driven-dissipative cold-atomic systems \cite{PhysRevLett.96.230602,PhysRevLett.99.140402,sieberer2016keldysh}. 
One interesting point is that some models show 
the time-dependent condensate, which can be regarded 
as the spontaneous symmetry breaking of time-translational symmetry 
in a \textit{strong} sense (See footnote \ref{footnote4} for our 
\textit{weak} and \textit{strong} criterion for SSB). 
However, we note that if we have e.g. the oscillating condensate 
associated with $U(1)$ symmetry, 
that state remains symmetric under the combination of 
time-translation and global $U(1)$ transformation.
Then, we can regard that symmetry breaking in terms of $U(1)$ symmetry breaking 
or the time-translation symmetry breaking. 
We thus need to clarify which description is better way 
to describe such systems.

\medskip
Also, there is a possibility to apply our formalism to chemical or biological 
systems. 
Indeed, there are a lot of open nonequilibrium systems 
such as the Belousov-Zhabotinsky reaction that shows synchronization phenomena. 
Again, this can be regarded as the time-translational symmetry breaking.
In order to derive the slow, or low-energy dynamics of systems, 
the so-called singular perturbation method has been traditionally used~\cite{Kuramoto,MoriKuramoto}.  
Although our formulation based on the effective Lagrangian and 
the singular perturbation method looks different, 
the basic philosophy to focus on the phase dynamics is shared.
Therefore, it may be interesting not only to apply our formalism 
but also to see the relation with the conventional method to treat 
the synchronization phenomena.

\acknowledgments 
The authors thank Y. Hidaka, Y. Minami and Pak Hang Chris Lau for useful discussions.
M.H. was supported by the Special Postdoctoral Researchers Program
at RIKEN.
S.K. is supported in part by the Senshu Scholarship Foundation.
T.N.  is in part supported by JSPS KAKENHI Grant Numbers JP17H02894 and JP18K13539, and MEXT KAKENHI Grant Number JP18H04352.
A.O. is supported by JSPS Overseas Research Fellowships.
This work was partially supported by the RIKEN iTHEMS Program 
(in particular, iTHEMS STAMP working group).

\appendix
\section{Derivation of mixing terms from environment}\label{kim}

In this appendix, starting from the microscopic total Lagrangian 
\eqref{eq:TotalL} with \eqref{eq:EachL}, 
we review how to derive mixing terms in Eq.~\eqref{eq:MSR3}, which represent
fluctuation and dissipation originated from couplings with 
environments~(See, e.g., \cite{Kamenev:2009jj} for a detailed discussion). 
For notational simplicity, we rescale $X\to M^{-1/2}X$ and $x_n\to m_n^{-1/2}x_n$, and correspondingly $g_n\to m_n^{1/2}M^{1/2}g_n$, to use canonically normalized variables throughout this section.
We first consider the simplest situation where the environment is composed of 
one harmonic oscillator, and later generalize the discussion 
into multi oscillator situation.
For that purpose, we here assume that the environment is thermalized 
at initial time $t_0$.

\begin{figure}[t]
\begin{center}
\includegraphics[width=15cm]{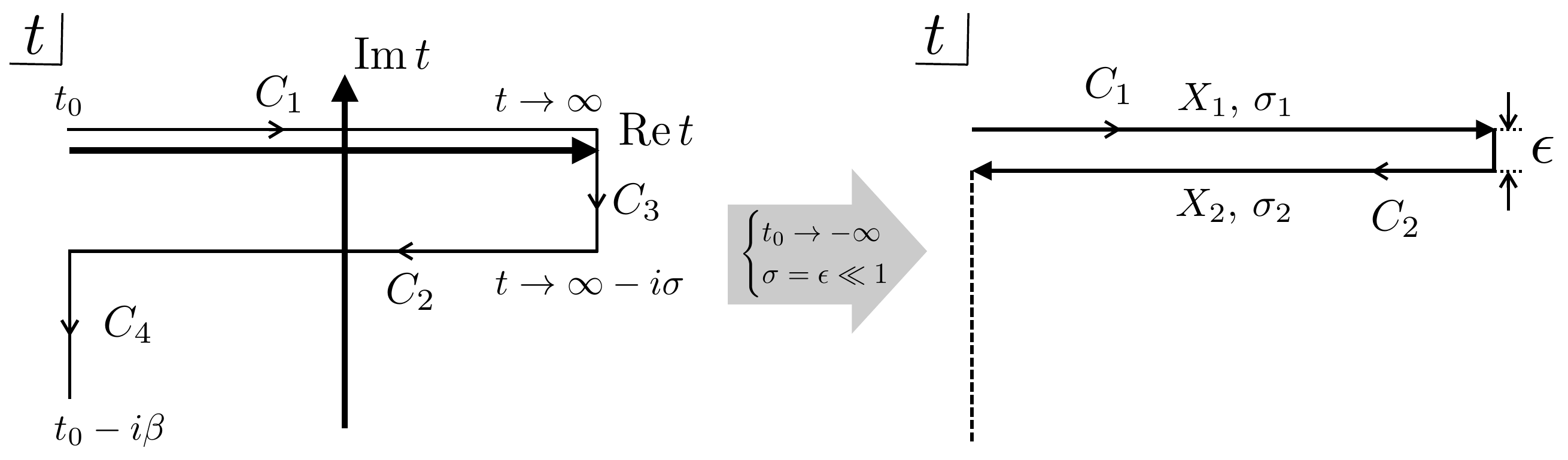}
\caption{
 Starting from the general closed-time-path (CTP) contour $C$ 
 for the initially thermalized system~(left figure),
 we choose the limit (right figure) in which the initial time is taken as 
 the past infinity ($t_0 \to - \infty$) and the forward and backward paths 
 are laid to overlap each other ($\sigma = \epsilon \ll 1$).
 In this limit, we only need to consider the  Green function 
 between the variables on the forward and the backward paths 
 expressed by $(X_1,\sigma_1)$ and $(X_2,\sigma_2)$, respectively.
 }
\label{app:ctp}
\end{center}
\end{figure}

\medskip
As is usual for the Schwinger-Keldysh formalism~\cite{Kamenev:2009jj,Bellac:2011kqa}, we first introduce a function $t = z(v)$ which parametrizes 
the closed-time-path (CTP) contour 
$C = \bigcup_a C_{a}~ (a=1,\cdots,4) $.
Here $v$ is taken as a monotonically increasing real parameter and 
$\sigma$ denotes a parameter which determines the imaginary-time 
position of the backward path $C_2$  (See the left in Fig.~\ref{app:ctp}).
Then, we define the step function and the $\delta$ function 
on the CTP contour $C$ as 
\begin{align}
 \theta_{C}(t-t') &\equiv \theta( v-v' ), \\
 \delta_{C} (t-t') 
 &\equiv \frac{d}{dt}\theta_{C}(t -t') 
 = \left( \frac{dz}{dv} \right)^{-1} \delta(v-v'),
\end{align}
where $\theta (v-v') $ and $\delta (t-t')$ are the usual step function 
and $\delta$ function.
With the help of these, we introduce the $2$-point real-time 
Green functions between the environment 
oscillator as
\begin{equation}
 \begin{split}
  G_C (t- t') 
  &\equiv 
  \average{T_C \big( \hx (t) \hx (t') \big)} \\ 
  &= \theta_C (t-t') \average{\hx (t) \hx (t')}
  + \theta_C (t'-t) \average{\hx (t') \hx (t)} \\
  &= \theta_C (t-t') G^> (t- t') 
  + \theta_C (t'-t) G^< (t- t'), 
 \end{split}
 \label{eq:CTPGreen}
\end{equation}
where $T_C$ denotes the time-ordered product on the CTP contour $C$, and
the angle bracket for an arbitrary operator $\hOcal$ does the thermal average:
\begin{equation}
 \average{\hOcal} \equiv 
  \mathrm{Tr} \big(\hat{\rho}_{\mathrm{eq}} \hOcal \big)  
  \with
  \hat{\rho}_{\mathrm{eq}} 
  \equiv \frac{1}{Z} e^{-\beta \hat{H}_{\mathrm{env}}} 
  ~\mathrm{and}~~
 \hat{H}_{\mathrm{env}} \equiv 
  \frac{\hp^2}{2} + \frac{1}{2} \omega_0^2 \hx^2 ,
\end{equation}
and $\beta$ being the inverse temperature of the environment.
In the last line of Eq.~\eqref{eq:CTPGreen}, 
we introduced the following greater and lesser Green functions:
\begin{equation}
  G^> (t- t') \equiv \average{\hx (t) \hx (t')}, \quad 
   G^<(t- t') \equiv  \average{\hx (t') \hx (t)} ,
\end{equation}
which, due to the initial thermal ensemble, 
satisfy the KMS (Kubo-Martin-Schwinger) condition:
\begin{equation}
 G^> (t- t') = G^< (t- t'+i\beta).
\end{equation}
Thanks to the KMS condition, 
the greater/lesser Green functions in the Fourier space can be expressed as
\begin{equation}
 G^> (\omega) = \big( 1 + n_B (\omega) \big) \rho (\omega), \quad  
  G^< (\omega) = n_B (\omega) \rho (\omega),
\end{equation}
where we introduced the spectral function $\rho (\omega)$ 
and the Bose-Einstein distribution $n_B (\omega)$ as follows: 
\begin{equation}
 \rho (\omega) \equiv G^>(\omega) - G^< (\omega), 
  \quad 
  n_B (\omega) \equiv \frac{1}{e^{\beta \omega} - 1}.
\end{equation}
In the following calculation, we will use the concrete form of the spectral function for 
the harmonic oscillator 
$\rho (\omega) = 2 \pi\,\mathrm{sgn} (\omega) \delta (\omega^2-\omega_0^2)$ 
with $\mathrm{sgn} (x)$ being the sign function.

\medskip
To obtain the effective action for the system 
$S_{\mathrm{eff}} [X_1,X_2]$, 
we first take our initial time to 
the past infinity $t_0 \to -\infty$ and choose the parameter 
$\sigma = \epsilon \ll 1$~(See the right figure in Fig.~\ref{app:ctp}). 
This considerably simplifies our problem 
because we only need to consider the correlations between variables on 
the forward and backward contours $C_{12} \equiv C_1 \cup C_2$. Since our action only contains terms linear and quadratic in the environment $x$, the effective action $S_{\mathrm{eff}} [X_1,X_2]$ may easily be calculated as
\begin{align}
\nonumber 
& iS_{\mathrm{eff}}[X_1,X_2] 
= i \int_{-\infty}^\infty dt \big(L_{\mathrm{sys}} (X_1) -L_{\mathrm{sys}} (X_2)\big)
\\
& \quad\quad - \frac{g^2}{2} \int_{-\infty}^\infty dt \int_{-\infty}^\infty dt' 
  \left(\begin{array}{cc}X_1(t) & -X_2(t)\end{array}\right)\left(\begin{array}{cc}G_{11} (t-t') & G_{12} (t-t') \\G_{21} (t-t') & G_{22} (t-t')\end{array}\right)\left(\begin{array}{c}X_1(t') \\-X_2(t')\end{array}\right).
  \label{eq:Seff}
\end{align}
where we introduced the following set 
of Green functions:
\begin{align}
\begin{split}
 G_{11}(t- t') &= \theta(t-t') G^> (t-t') + \theta(t'-t) G^< (t-t'),
 \\
 G_{12}(t- t') &= G^{<}(t-t'),
 \\
 G_{21}(t- t') &= G^{>}(t-t'),
 \\
 G_{22}(t- t') &= \theta(t'-t) G^>(t-t') + \theta(t-t') G^<(t-t').
\end{split}
\end{align}
Then, using 
$\rho (\omega) = 2 \pi\,\mathrm{sgn} (\omega) \delta (\omega^2-\omega_0^2)$,
we obtain an explicit form of all Green functions in the Fourier space 
as 
\begin{align}
\begin{split}
 G_{11} (\omega;\omega_0^2)
 &=  \frac{i}{\omega^2-\omega_0^2 + i \epsilon} 
 + 2 \pi n_B (|\omega|) \delta (\omega^2-\omega_0^2)\\
 &=\mathbf{P}\left(\frac{i}{\omega^{2}-\omega_{0}^{2}}\right)
+2\pi{\rm sgn}(\omega)\delta(\omega^{2}-\omega_{0}^{2})\left(\frac{1}{2}+n_B(\omega)\right)  ,
 \\
 G_{12} (\omega;\omega_0^2) 
 &= 2 \pi\,\mathrm{sgn} (\omega) \delta (\omega^2-\omega_0^2) n_B (\omega) ,
 \\
 G_{21} (\omega;\omega_0^2) 
 &= 2 \pi\,\mathrm{sgn} (\omega) \delta (\omega^2-\omega_0^2) 
 \big( 1 + n_B (\omega) \big),
 \\
 G_{22} (\omega;\omega_0^2) &= \big( G_{11} (\omega;\omega_0) \big)^* ,
 \end{split}
 \label{def:Greenfunc}
\end{align}
where $\mathbf{P}$ denotes a principal value and we explicitly wrote the $\omega_0$-dependence for later purpose.
These expressions enable
us to obtain the mixing terms between $X_1$ and $X_2$ in 
the effective action for the system \eqref{eq:Seff}.
Note that they appear as a direct consequence of the integrating out, 
or coarse-graining procedure of the environment. 

\medskip
We finally generalize our single harmonic oscillator 
result~\eqref{def:Greenfunc} to the multi harmonic oscillator case.
It can be easily performed by the replacement,
\begin{align}
 g^2G_{ab} (\omega;\omega_0^2) 
  \to \mathcal{G}_{ab} (\omega)
  \equiv \sum_{n=1}^N g_n^2G_{ab} (\omega;\omega_n^2)\,,
\end{align}
which characterizes the environment effects on the Brownian particle dynamics at each scale $\omega$. Especially when we are interested in the large $N$ limit, or in other words the continuous spectrum, it is convenient to introduce a weight function $J(\omega_0^2)$ such that
\begin{align}
 \mathcal{G}_{ab} (\omega)
  \equiv \frac{1}{2\pi} \int_{0}^{\Lambda^{2}} d\omega_0^{2} 
 J(\omega_0^2) G_{ab} (\omega;\omega_0^2), 
 \label{eq:GreenMulticase} 
\end{align}
where we introduced a cutoff scale $\Lambda$ for the environment distribution. Since the harmonic oscillator with a high frequency $\omega_0\gg T$ is not thermally excited very much, its effect on the Brownian particle dynamics will be negligible. Let us therefore assume that $\Lambda\sim T$.

\medskip
From now on, let us focus on the small frequency range compared to the temperature scale $\omega\ll T$ and suppose that the weight function takes the form,
\begin{equation}
\label{Ohmic}
 \mathrm{sgn}\, (\omega) J(\omega^2) 
  = 2\gamma\omega\Big[ 1 + O \left(\frac{\omega^2}{\Lambda^2}\right) \Big],
\end{equation}
where note that $ \mathrm{sgn}\, (\omega) J(\omega^2) $ is an odd function of $\omega$, hence the leading order is a linear term as long as it is finite at $\omega=0$. We set the coefficient of the linear term to be $2\gamma$ ($>0$), which has a mass dimension one (the factor $2$ is for later convenience).
The environment described by this weight function is known 
as the \textit{Ohmic bath}~\cite{kamenev2011field}.
In this case, we can perform $\omega_0^2$ integration 
in Eq.~(\ref{eq:GreenMulticase}) and obtain
\begin{equation}
 \begin{split}
\mathcal{G}_{11} (\omega)
  &\simeq \gamma(-iC
  + 2T)\,,
\\
  \mathcal{G}_{12} (\omega) 
  &\simeq \gamma(-\omega 
  + 2T)\,,
\\
  \mathcal{G}_{21} (\omega)
  &\simeq  \gamma(\omega
  + 2T)\,,
\\
  \mathcal{G}_{22} (\omega) &\simeq\gamma(iC
  + 2T)\,,
\end{split} 
\end{equation}
where we introduced $C=2\Lambda/\pi$ and dropped higher-order terms in the $\omega/T$ expansion. In the real-time coordinate, we have
\begin{equation}
 \begin{split}
\mathcal{G}_{11} (t,t') &\simeq \gamma\delta(t-t')\left(- iC 
  + 2T\right),
\\
\mathcal{G}_{12} (t,t')& \simeq \gamma\delta(t-t')\left(- i \partial_{t'}
  + 2T\right),
 \\
 \mathcal{G}_{21} (t,t') 
 &\simeq \gamma\delta(t-t')\left( i\partial_{t'}
  + 2T\right),
 \\
 \mathcal{G}_{22} (t,t') 
 &\simeq \gamma\delta(t-t')\left( iC 
  + 2T\right).
\end{split} 
\end{equation}
We eventually obtain the effective action \eqref{eq:Seff} as 
\begin{align}
\nonumber
 iS_{\mathrm{eff}}[X_1,X_2]
&
 =   i \int_{-\infty}^\infty dt \big(L_{\mathrm{sys}} (X_1) -L_{\mathrm{sys}} (X_2)\big)
 \\
&\quad  -\frac{1}{2}
  \int_{-\infty}^\infty dt 
 \left[ i\gamma\left(X_1\dot{X}_2-\dot{X}_1X_2\right) +2\gamma T\left(X_1^2+X_2^2-2X_1X_2\right)\right],
\end{align}
where we absorbed the constant $C$ into the potential $V(X)$ of the Brownian particle by renormalization.
This gives the effective action~\eqref{eq:MSR3} after the rescaling $X\to M^{1/2}X$.

\bibliography{bib}{}
\bibliographystyle{unsrt}

\end{document}